\documentclass[preprint,11pt]{aastex}

\newcommand{\degree}{\mbox{$^{\circ}$}}
\newcommand{\am}{\mbox{\arcmin}}
\newcommand{\as}{\mbox{\arcsec}}

\def\lsim {$\rlap{\raise.4ex\hbox{$<$}}\lower.55ex\hbox{$\sim$}\,$}

\newcommand{\lsun}{\mbox{L$_\odot$}}

\input{epsf}

\def\deg{{$^\circ$}}

\begin{document}


\title{\bf The {\it Spitzer} c2d Survey of Nearby Dense Cores: III: Low Mass 
Star Formation in a Small Group, L1251B}
\author {Jeong-Eun Lee\altaffilmark{1,5},
James Di Francesco\altaffilmark{2},
Shih-Ping Lai\altaffilmark{3,8,9},
Tyler L. Bourke\altaffilmark{4},
Neal J. Evans II\altaffilmark{5},
Bill Spiesman\altaffilmark{5},
Philip C. Myers\altaffilmark{4},
Lori E. Allen\altaffilmark{4},
Timothy Y. Brooke\altaffilmark{6},
Alicia Porras\altaffilmark{4},
Zahed Wahhaj\altaffilmark{7}
}
\altaffiltext{1}{Hubble Fellow, Physics and Astronomy Department, The University
of California at Los Angeles, PAB, 430 Portola Plaza, Box 951547, Los Angeles, 
CA 90095-1547; jelee@astro.ucla.edu}
\altaffiltext{2}{National Research Council of Canada, Herzberg Institute of 
Astrophysics, 5071 West Saanich Road, Victoria, BC V9E 2E7, Canada; 
james.difrancesco@nrc-cnrc.gc.ca}
\altaffiltext{3}{Department of Astronomy, University of Maryland, College Park, 
MD 20742; slai@astro.umd.edu} 
\altaffiltext{4}{Smithsonian Astrophysical Observatory, 60 Garden Street, 
Cambridge, MA 02138; tbourke@cfa.harvard.edu, pmyers@cfa.harvard.edu,
leallen@cfa.harvard.edu, aporras@cfa.harvard.edu}
\altaffiltext{5}{Astronomy Department, The University of Texas at Austin,
1 University Station C1400, Austin, TX 78712-0259; 
nje@astro.as.utexas.edu, spies@astro.as.utexas.edu}
\altaffiltext{6}{Division of Physics, Mathematics, \& Astronomy, MS 105-24, 
California Institute of Technology, Pasadena, CA 91125; tyb@astro.caltech.edu}
\altaffiltext{7}{Northern Arizona University, Department of Physics and 
Astronomy, Box 6010, Flagstaff, AZ 86011-6010; zwahhaj@physics.nau.edu}
\altaffiltext{8}{Institute of Astronomy and Department of Physics, National
Tsing Hua University, Hsinchu 30043, Taiwan}
\altaffiltext{9}{Academia Sinica Institute of Astronomy and Astrophysics, P.O.
Box 23-141, Taipei 106, Taiwan}

 
\begin{abstract}
We present a comprehensive study of a low-mass star-forming region, 
L1251B, at wavelengths from the near-infrared to the millimeter.  L1251B, 
where only one protostar, IRAS 22376+7455, was known previously, 
is confirmed to be
a small group of protostars based on observations 
with the {\it Spitzer} Space Telescope.  The most luminous source of L1251B 
is located 5$\arcsec$ north of the IRAS position.  A near-infrared bipolar 
nebula, which is not associated with the brightest object and is located at the 
southeast corner of L1251B, has been detected in the IRAC bands.  
OVRO and SMA interferometric observations indicate that the brightest
source and the bipolar nebula source in the IRAC bands are deeply embedded
disk sources.
Submillimeter continuum observations with single-dish telescopes and the SMA 
interferometric observations suggest two possible prestellar objects with 
very high column densities.
Outside of the small group, many young stellar object candidates have been 
detected over a larger region of $12\arcmin \times 12\arcmin$.  Extended 
emission to the east of L1251B has been detected at 850 \micron; this ``east
core" may be a site for future star formation since no point source has been 
detected with IRAC or MIPS. 
This region is therefore a possible example of low-mass cluster 
formation, where a small group of pre- and protostellar objects (L1251B) 
is currently forming, alongside a large starless core (the east core). 
\end{abstract}


\section{INTRODUCTION}

Stars form out of dense cores of molecular gas, but the details of this
process are far from understood.  To probe this process, continuum and 
line data of star-forming regions can be combined to provide a coherent, 
self-consistent picture of young stellar objects and their associated,
gaseous environments, i.e., densities, temperatures, and velocities.  
Indeed, these physical properties are intertwined within dynamical and 
luminosity evolution during star formation.  Observations over a wide 
range of continuum wavelengths can test models of dynamical and luminosity 
evolution because different parts of a star-forming region are traced 
by different wavelengths.  For example, cold envelopes are traced by 
(sub)millimeter dust emission which yield information on total (gas+dust) 
masses.  On the other hand, inner envelopes or disks are traced by 
infrared emission, which is directly related to internal luminosity sources.  
In addition, observations of a wide variety of lines yield information 
about associated gas motions and can probe the excitation conditions 
along various lines of sight in star-forming regions.  Chemistry is also 
important to understand, however.  For example, velocity structures derived 
from molecular spectra can be misinterpreted if incongruous abundance 
profiles are adopted.  Several studies have shown that chemical evolution 
is not independent of dynamical and luminosity evolution (Rawlings \& 
Yates 2001; Doty et al. 2002; Lee, Bergin \& Evans 2004).  

Most star formation in the Galaxy occurs in clusters embedded within giant 
molecular clouds (Lada \& Lada 2003), but the high surface densities of 
young stellar objects make it difficult to disentangle the star formation 
process.  Although not considered ``typical" star formation, more isolated 
cases have been examined vigorously since they are much less complicated 
and observations of them are easier to interpret (Di Francesco et al. 2006).  
Between these extremes lie the smaller {\it groups} of young stellar objects 
(i.e., $N$ $<$ 10), which may have condensed out of a single dense core 
(Machida et al. 2005).  Examining such groups in detail 
may provide access to understanding features of the clustered star formation 
process, yet are relatively uncomplicated and hence easier to understand.
Indeed, our Galactic neighborhood is resplendent in examples of groups of
young stellar objects (YSOs), whose associated continuum and line emission 
can be examined to probe how they formed.

In this paper, we examine continuum emission associated with L1251B, 
where Spitzer Space Telescope (SST) observations reveal a small group.  
The line emission associated with L1251B is presented in a separate paper
(Lee et al. in preparation; hereafter Paper II).
L1251B is located at 300 $\pm$ 50 pc (Kun \& Prusti 1993), and lies within 
the densest of five C$^{18}$O cores observed by Sato et al. (1994).  Sato 
et al. (1994) named the core ``E", but Hodapp (1994) called it ``L1251B" 
because it is associated with the weaker of two outflow sources found by 
Sato \& Fukui (1989).  In this paper, we use ``L1251E" to refer to the more 
extended region observed in C$^{18}$O 1$-$0 and covering five IRAS point 
sources shown in Figure 3 of Sato et al. (1994).  (SST observations cover 
this extended region, and are thus named ``L1251E" in the {\it Spitzer} data 
archive.)   On the other hand, we use ``L1251B" to refer to the densest region 
around IRAS 22376+7455, which has been cited as the driving source of 
the molecular outflow in L1251B (Sato et al. 1994).  

Earlier observations have shown that star formation is proceeding in L1251B, 
but details have been difficult to discern, due to low sensitivity and 
resolution.  The $K'-$band image of L1251B by Hodapp (1994) shows a bipolar 
nebula with nearby infrared sources, giving a clue that a small group
may be forming.  Myers \& Mardones (1998) also found five ISO sources within 
L1251B, further suggesting the possibility of grouped star formation.  
We have gathered new observations of L1251B at high sensitivity or resolution, 
to probe in detail the formation and interaction processes in this small group. 
To determine which sources are YSOs within L1251B, we include new 
continuum data from near-infrared to millimeter wavelengths.
In this study, we use coordinates in the J2000 epoch, so the ($\alpha$, 
$\delta$) coordinates of IRAS 22376+7455 are 
(22$^{h}$38$^{m}$47.16$^{s}$, +75\deg 11\am 28.71\as).  

Observations of L1251B are summarized in \S 2.  Results and analysis are shown 
in \S 3 and \S 4 where we zoom in from large to small scales in the description 
of the results.  In \S 5, we compare L1251E and L1251B with other star forming
regions observed with the SST and discuss a possible scenario of the formation 
of L1251B within L1251E. 
Finally, a summary appears in \S 6.

\section{OBSERVATIONS}
\subsection{The {\it Spitzer} Space Telescope (SST) Observations}

The SST Legacy Program ``From Molecular Cores to Planet Forming Disks" (c2d; 
Evans et al.\ 2003), observed L1251E at 3.6 $\mu$m, 4.5 $\mu$m, 5.8 $\mu$m, 
and 8.0 $\mu$m with 
the Infrared Array Camera (IRAC; Fazio et al. 2004) on 29 October 2004 and at 
24 $\mu$m and 70 $\mu$m with the Multiband Imaging Photometer for Spitzer 
(MIPS; Rieke et al. 2004) on 24 September 2004 (PID:139;  AOR keys: IRAC 5167360
and MIPS 9424384). The diffraction limit is 1\farcs1 at 3.6 $\mu$m and 7\farcs1 
at 24 $\mu$m for the 85 cm aperture, and the pixel sizes are 1\farcs2 for 
all IRAC bands and 2\farcs55 at MIPS 24 $\mu$m.
 
The L1251E IRAC map contains 20 pointings (4$\times$5 field of views, 
each $5\am \times 5\am$ in area).  Four dithers were performed at each 
pointing position and the exposure times were 12 s for each dither, 
resulting in 48 s exposure time for most of the pixels.  These images were 
combined in each band to form $\sim$0\fdg4\ $\times$ 0\fdg35\ mosaic images 
centered at ($\alpha$, $\delta$) = (22$^{h}$38$^{m}$15.32$^{s}$, 
+75\deg 8\am 46.68\as) for the 3.6 $\mu$m and 5.8 $\mu$m bands and 
(22$^{h}$39$^{m}$20.52$^{s}$, +75\deg 14\am 9.24\as) 
for the 4.5 $\mu$m and 8.0 
$\mu$m bands.  MIPS images were obtained with photometry mode over 9 pointings. 
With dithering, the total observed area is $\sim$0\fdg3\ $\times$ 0\fdg3 at 
24 $\mu$m centered at ($\alpha$, $\delta$) = 
(22$^{h}$38$^{m}$47.23$^{s}$, +75\deg 11\am 29.39\as) and the 
exposure time is 48.2 s for most of the pixels.  

The IRAC and MIPS images were processed by the {\it Spitzer} Science Center 
(SSC) using the standard pipeline (version S11) to produce Basic Calibrated Data
(BCD) images.  The BCD images were further processed by the c2d team to improve 
their quality.  The c2d processed images were used to create 
improved source catalogs, and both improved images and catalogs were delivered 
to the SSC Data Archive.  

A detailed discussion of the data processing and products can be found in 
the c2d delivery documentation (see Evans et al. 2006), but here we give a 
brief overview.  First, the BCD images were examined, an additional bad 
pixel mask was created, and additional corrections were done for muxbleed, 
column pulldown, banding, and jailbar features.  Next, the resulting 
post-BCD images were mosaicked using the SSC Mopex tool with outlier 
rejection and position refinement.  Source extraction was done on these 
images utilizing a specialized version of ``Dophot" developed by the c2d 
team.  The source lists for each band were compared with the images to 
remove false sources caused by diffraction spikes, saturation, and other 
imperfections.  The purified source list at each band was merged with each 
other as well as the 2MASS catalog to create the final catalog; position 
coincidence within 4\arcsec\ of the MIPS detection and 2\arcsec\ of the 
detections in all other bands are cataloged as the same source.  

\subsection{The Caltech Owens Valley Radio Observatory (OVRO) Millimeter 
Array Observations}

L1251B was observed in the 
1 mm and 3 mm bands of the OVRO Millimeter Array near Big Pine, CA over 
two observing seasons, Fall 1997 and Spring 1988.  
Continuum emission at $\lambda$ = 1.33 mm and 2.95 mm were observed over 
2 L-configuration and 1 H-configuration tracks for spatial frequency coverages 
of $\sim$11-156 k$\lambda$ for the $\lambda$ = 1.33 mm data 
and $\sim$5-71 k$\lambda$ for the $\lambda$ = 2.95 mm data.  
These data were retrieved from the OVRO archive.  
Table 1 summarizes these data, i.e., the wavelengths observed and the band 
widths used, as well as 
the synthesized beam FWHMs and the 1 $\sigma$ rms values achieved.  
The radio sources 1928+738 or BL Lac were 
observed for $\sim$5 minutes at intervals of $\sim$20 minutes between 
visits to L1251B to provide gain calibration.  3C 84, 3C 273, 
3C 345, 3C 454.3, Uranus, or Neptune were observed at the beginning or 
end of each track for $\sim$5--10 minutes to provide flux and passband 
calibration.

All data were calibrated using standard procedures of the Caltech MMA 
package (Scoville et al. 1993).  The flux calibration is likely accurate 
to within $\sim$20\%.  Calibrated L1251B and gain calibrator visibilities 
were then converted to FITS format and maps were made using standard tasks 
of the MIRIAD software package (Sault, Teuben \& Wright 1995).  Table 1 
also lists the FWHMs of Gaussians applied to the visibility data during 
inversion to the image plane to improve sensitivity to extended structure.  
All data were deconvolved using a hybrid Clark/H\"ogbom/Steer CLEAN 
algorithm to respective levels of $\sim 2$ $\sigma$ per channel.

\subsection{The Submillimeter Array (SMA) Observations}

L1251B was observed in the 1.3 mm continuum on 25 September 2005 with
the Submillimeter Array (SMA; Ho, Moran \& Lo 2004\footnote{The
Submillimeter Array is a joint project between the Smithsonian Astrophysical
Observatory and the Academia Sinica Institute of Astronomy and Astrophysics 
and is funded by the Smithsonian Institution and the Academia Sinica.}) with 6
antennas in a compact configuration.  The observations covered the spatial
frequency range of $\sim5.5 - 42$ k$\lambda$.  Two overlapping positions
centered on IRS1 and IRS2, separated by 22\arcsec, were observed
alternatively, and were interleaved with observations of the quasars
J2005+778 and J0102+584 for gain calibration.  The quasar 3C454.3 was
observed for passband calibration, and Uranus was observed for flux
calibration.

Calibration was performed with the IDL-based MIR software package (Qi
2006) following standard procedures.  The SMA is a double-sideband instrument,
with each sideband covering 2 GHz, and separated by 10 GHz.  Data in the
upper and lower sidebands were calibrated in MIR separately, and the
continuum was derived from line-free channels in each sideband.  Imaging
was performed in MIRIAD (Sault, Teuben \& Wright 1995).  The continuum
data from each sideband were combined during inversion to improve the
sensitivity, and the effective frequency of the combined continuum data
is 225.404 GHz (1.33 mm). Natural weighting was used without any tapering during
inversion. The two L1251B pointings were imaged as a mosaic
with appropriate primary beam correction (the SMA primary beam is
$\sim55\arcsec$ at this frequency), and deconvolved using a
Steer-Dewdney-Ito CLEAN algorithm optimized for mosaic observations
(MIRIAD task mossdi2), down to the respective 1$\sigma$ levels in the dirty
image.
The observation is summarized in Table 1.

\subsection{Data from Other Observations} 

An 850 $\mu$m continuum emission map covering the L1251E region was obtained 
from the James Clerk Maxwell Telescope (JCMT) Submillimeter Common User 
Bolometer Array (SCUBA) mapping database compiled by Di Francesco et al. (in 
preparation).  This map consists of data obtained from the JCMT public archive 
maintained by the Canadian Astronomical Data Centre\footnote{The Canadian 
Astronomical Data Centre is operated by the Dominion Astrophysical Observatory 
for the National Research Council of Canada.} (CADC).  The data were processed 
using a matrix inversion technique (see Johnstone et al. 2000) and the most 
current determinations of extinction and flux conversion factors.  The maps 
were flattened and edge-clipped to remove artifacts and smoothed with a 
Gaussian of $\sim$14\as\ FWHM to reduce pixel-to-pixel noise.  The final 
resolution of the map is $\sim$20\as\ FWHM.  (C. Young also provided SCUBA 
data of L1251B alone at 450 $\mu$m and 850 $\mu$m that were originally 
published by Young et al. 2003.)

The c2d program also observed L1251B at 350 $\micron$ (Wu et al., in 
preparation).  The 350 $\mu$m continuum observations of L1251B were acquired in 
May 2005, using the SHARCII instrument on the Caltech Submillimeter Observatory
(CSO).  SHARCII is equipped with 
a 12 $\times$ 32 pixel array, resulting in a 2\farcm59 $\times$ 0\farcm97
field of view.  The beam size with good focus and pointing is about 8$\as$ 
at 350 $\mu$m. 

M. Tafalla provided an unpublished 1.3 mm dust continuum map of L1251B, 
observed with the MAMBO instrument on the IRAM 30-m Telescope at Pico 
Veleta, Spain, as part of a larger survey carried out in February 1997
(see Tafalla et al. 1999 for observing details). 

\section{OBSERVATIONAL RESULTS}

Figure 1 shows the Digitized Sky Survey (DSS) R-band image around L1251E, with 
stars identified as ``H$\alpha$ stars" by Kun \& Prusti (1993) indicated. 
In comparison, Figure 2 shows the IRAC three-color image of 3.6, 4.5, and 8.0 
\micron, revealing a small group of infrared sources associated with L1251B, 
at the location of one H$\alpha$ star and adjacent to the position of IRAS 
22376+7455.  
The spiral galaxy, UGC 12160 (Cotton et al. 1999), is located in the east edge
of Figure 2.
To the northwest, Figure 2 also shows the Herbig-Haro (HH) object
HH373, which shows up in the longer wavelength bands as well as at 4.5 $\mu$m.  
The extended green color seen in Figure 2 all around the central group is 
dominated by scattered light with some contribution from shocked gas associated 
with jets and outflows.  A jet-like feature, $\sim$0.3 pc long and aligned in 
the SE-NW direction, is seen centered near the IRAS position, with an emission 
knot at its northwestern tip.  There is also green, arc-shaped emission 
$\sim$2\arcmin\ ($\sim$0.17 pc) south of L1251B in Figure 2, which is 
associated with HH189C (Eiroa et al. 1994) and might be caused by a jet from 
one protostar of the group.  There are red-free green patches, 
indicating absorption 
against the background 8.0 \micron\ PAH emission, to the east of the central 
group.  These patches are dominated by scattered light and seen as optically 
dark locations in the DSS R-band 
image (Figure 1) and as submillimeter bright locations in the 850 \micron\ 
emission map (see Figure 3).

In Figure 3, extended emission at 850 \micron\ in the L1251E region is shown
with contours on top of the MIPS 24 $\micron$ image.  The 850 $\micron$ 
emission peaks very strongly at L1251B, which we call the ``west core."  
There are 3-4 other, weaker, 850 $\mu$m peaks within 200$\as$ of L1251B 
to the east (NE to ESE).  The brightest of these, which we call the ``east 
core", is located $\sim$180\as\ east of L1251B.  Three 24 \micron\ sources 
are located east of L1251B region and two are coincident with T Tauri-like 
objects with H$\alpha$ emission (Kun \& Prusti 1993).  
There are no sources, however, in the IRAC and MIPS 
bands around the weaker 850 $\mu$m emission peaks.  T$\acute{\rm o}$th \& 
Walmsley (1996) showed that the east core has a higher column density of 
NH$_3$ than the west core (L1251B) and that both may be gravitationally 
unstable.  Therefore, the east core may be prestellar in nature, i.e., a 
potential site of future star formation.  
Only three (two T Tauri-like objects and IRAS 22376+7455) of five IRAS point 
sources have corresponding objects at 24 $\micron$ in Figure 3.  
The other two IRAS point sources have not been detected either in IRAC or MIPS 
bands.

Figure 4 shows the zoomed-in three-color composite image of L1251B, the 
central group of Figure 2, while Figure 5 shows separately the IRAC 3.6 
\micron, 4.5 \micron, and the 5.8 \micron\ band images and the MIPS 24 
\micron\ band image around L1251B.
There are 8 point-like sources in the central region of Figure 4.  The source 
denoted with a triangle at (22$^{h}$38$^{m}$49.7$^{s}$, +75\deg 11\am 52.7\as) 
is detected as a point source only in the IRAC 3.6 and 4.5 \micron\ bands, 
however.  The red source denoted with a square at (22$^{h}$38$^{m}$38.7$^{s}$, 
+75\deg 11\am 43.9\as) is weak and nebulous in the short IRAC bands, but in 
the IRAC 8.0 \micron\ band, the source becomes stronger and point-like on top 
of arc-shaped emission.  This source might be an infrared source analog to a 
HH object.  Therefore, there 
are 6 bright sources detected in all IRAC bands as point sources, and among 
them, only the three sources, each marked with an X, are clearly identified 
at 24 \micron.  IRAS 22376+7455 covers the whole group because of the poor 
resolution of the IRAS observations.  The coordinates of the IRAS source, 
however, are closest to the brightest IRAC source, which we name 
``L1251B-IRS1."  
A bipolar structure associated with the SE source, which we name ``L1251B-IRS2,"
is clearly seen, and it is probably due to scattered light from a bipolar 
outflow cavity.  The opening of the bipolar nebula is similar in direction to 
that of the jet seen (with the white line) in Figure 2, suggesting that 
IRAS 22376+7455 may not be 
the only driver of the large CO outflow, as previously believed.  (Note that
the bipolar nebula is {\it not} associated with IRS1.)  We will discuss 
this situation further in Paper II.

Figure 6 shows the continuum emission around L1251B at various (sub)millimeter 
wavelengths on top of the IRAC 4.5 \micron\ band image.  The IRAC image shows 
the bipolar nebula of IRS2 and diffuse emission over the region likely 
due to scattering.  There are two intensity peaks at 350 $\micron$ and 450 
$\micron$; the stronger peak does not have a corresponding IRAC source and 
the weaker peak seems related to IRS1, the brightest IRAC source.  The 
emission at 850 $\micron$ and 1300 $\micron$, however, has only one peak 
corresponding to the stronger peak at 350 $\micron$ and 450 $\micron$.  The 
stronger intensity peak is located between IRS1 and IRS2.  (As we show 
in Paper II, this position is coincident with a peak of N$_2$H$^+$ 
emission, which traces cold, dense regions.)  The stronger peak may be a high 
column density peak of the envelope.  The weaker peak, however, may be heated 
primarily by IRS1 since it disappears at longer wavelengths.

If the intensity distributions of dust continuum at various wavelengths are
compared, a shift is seen in the positions of intensity peaks with wavelength 
from north to south.  The pointing error of the 350 $\micron$ emission is 
relatively large, so it could be adjusted in position to match its weaker 
peak to the 450 $\micron$ second peak.  A systematic shift of intensity peaks 
with wavelength, however, cannot be explained only by pointing errors.  The 
angular separation ($\sim6\as$) between the 1.3 mm intensity peak and the 
stronger 450 $\micron$ peak is much greater than the pointing error ($<3\as$) 
of the 450 $\micron$ and 1.3 mm observations.  Such a shift may involve a 
systematic variation of physical conditions such as density, temperature, or 
opacity. 

Figure 7a shows the dust continuum intensities observed with the OVRO array 
at 3 mm and 1.3 mm.  Such data can trace dust from deeply embedded disks or 
the interiors of envelopes.  The two intensity peaks at 3 mm correspond to 
IRS1 and IRS2, as seen in Figure 7b.  At 1.3 mm, however, IRS2 is 
located beyond of the FWHM of the primary beam at the coordinates of IRAS 
22376+7455.  The OVRO results also show that IRS1 is located 5$\as$ north 
of the coordinates given in the IRAS point source catalog.

The result of the SMA observation in the 1.3 mm continuum is shown in Figure 8.
Four detections are seen.  
The brighter two detections are compact and coincident with IRS1 and IRS2 
as seen in the OVRO 3 mm data, so deeply embedded disk components of IRS1
and IRS2 are traced with these interferometric observations.
The fainter two detections are more extended and peak to the east and southeast 
of IRS1, and they were detected separately in the USB 
and LSB data sets which were reduced independently. 
The faint extended emission associated with these two latter sources is 
coincident with the brightest regions of integrated N$_{2}$H$^{+}$ 1$-$0 
intensity seen in our OVRO map (see Paper II).  
No 1.3 mm continuum
emission is seen associated with any other near-infrared sources.
The two weaker intensity peaks at 1.3 mm are also consistent with the stronger
peaks of the 350 and 450 \micron\ emission and the peaks of 830 and 1300 
\micron\ emission observed with single dishes. 
Therefore, the two intensity peaks newly detected in the SMA 1.3 mm observation 
are probably the highest densities of the L1251B envelope, whose extended 
emission was resolved out by the interferometers.
If true, these sources may be prestellar
condensations and not detected at 3 mm with OVRO due to low sensitivity.

Interestingly, however, the two weaker SMA 1.3 mm emission peaks were not 
detected in the OVRO 1.3 mm observation. The two SMA 1.3 mm peaks lie within
the FWHM of the OVRO primary beam, and the 1 $\sigma$ sensitivities of both 
OVRO and SMA 1.3 mm observations are the same although the peak intensity of 
IRS1 in the OVRO 1.3 mm observation (14 mJy beam$^{-1}$) is less than a half
of that in the SMA 1.3 mm observation (34 mJy beam$^{-1}$).   
We tested the SMA data by excluding spatial frequencies lower than the
lowest spatial frequency of the OVRO data, and found that extended 1.3 mm 
emission between IRS1 and IRS2 disappeared in the resulting images.
This discrepancy is therefore caused mainly by shorter baselines in
the SMA data ensemble, which is more sensitive to extended structure.
Remaining discrepancies are likely the result of the flux calibration
not being as accurate as for the SMA data since flux calibrators were
monitored at 1 mm less frequently at OVRO.  As a result, the flux of
IRS1 from the SMA observation is stronger than that from the OVRO
observation.  Other complementary data, such as continuum data at 345 GHz 
(observed with the SMA), are necessary to understand the physical
conditions of two extended prestellar condensations.

\section{ANALYSIS}

For general analysis with color-color and color-magnitude diagrams, we include 
all sources in Figure 2, but focus on sources around L1251B for more detailed 
analysis of their spectral energy distributions (SEDs).  Figure 9 shows our
selected identification numbers for sources around L1251B. 
Here, 6 sources (from 1 to 6) within the L1251B dense core, which is detected 
in submillimeter continuum with single-dish telescopes (Figure 6) and focused on
in Paper II, are named as L1251B-IRS1, -IRS2, -IRS3, and so on. 
Again, IRS1 is the brightest IRAC source, 
and IRS2 is the object associated with the NIR bipolar nebula.  
Other sources (from 7 to 16) outside L1251B are just named as source 7, 
source 8, and so on. 

Allen et al. (2004) described how the [3.6]$-$[4.5] vs.\/ [5.8]$-$[8.0] 
color-color diagram can be used to identify the evolutionary stages of 
sources detected by IRAC.  Figure 10 shows such a color-color diagram for 
sources in L1251E, along with domains of color representative of objects 
from various pre-main-sequence classes, similar to those identified by 
Allen et al.  For example, we identify Class II sources in L1251B with 
the color range 0 $<$ [3.6]$-$[4.5] $<$ 0.8 and 0.4 $<$ [5.8]$-$[8.0] $<$
1.1, and Class 0/I sources within color range of [3.6]$-$[4.5] $>$ 0.8 and 
0.4 $<$ [5.8]$-$[8.0] $<$ 1.5.  This latter range covers a smaller area 
than that of the Class 0/I models shown in Figure 4 of Allen et al., since
none of the sources with [5.8]$-$[8.0] $>$ 1.5 are in the L1251B core where 
Class 0/I sources are most likely located.  According to these color 
ranges, there are 33 Class II sources and 10 Class 0/I sources for L1251E 
as a whole. 

Background galaxies can be detected with the sensitivity of the c2d 
observations, and may have colors within the expected ranges of Class 0/I 
or Class II sources (Eisenhardt et al. 2004).  To remove potential galaxy 
``contamination," we require the YSO candidates fulfill either one of two 
sets of color-magnitude criteria: (1) [4.5]$-$[8.0] $>$ 0.5 and [8.0] $<$ 
(14 $-$ [4.5]$-$[8.0]) {\it or} (2) [8.0]$-$[24] $>$ 0.7 and [24] $<$ (12 
$-$ [24]$-$[8.0]).  These criteria are derived from a comparison 
(Harvey et al. 2006; Jorgensen et al. 2006) between 
the c2d data and the data from the SST SWIRE Legacy Project.
The SWIRE project is a deep imaging survey 
at the Galactic Pole and provides an essentially YSO-free sample, while 
the regions observed by the c2d project contain a fair amount of YSOs.  The 
comparison between the two data sets show that the SWIRE sources are 
extremely depleted in the color-magnitude space defined by the two sets 
of color-magnitude criteria. 
Two sets of color-criteria are shown in Figure 11 with dashed lines.
The level of the criteria are chosen so that (1) 95\% of the SWIRE
sources with [4.5]$-$[8.0] $>$ 0.5 are under the diagonal line in the left
panel and (2) 95\% of the SWIRE sources with [8]$-$[24] $>$ 0.7 are under the
diagonal line in the right panel. Under these criteria only 15 SWIRE
sources per square degree could be potentially mis-identified as YSO
candidates. (A detailed description of these selection criteria is
provided in Lai et al., in preparation). The probability of the faint
YSOs missed by these criteria cannot be assessed at this point due to
lack of knowledge of faint YSO populations.

Figure 11 shows the distribution of the sources detected in L1251E with peak 
signal-to-noise ratios $\ge$ 7 in all IRAC bands in [8.0] vs. [4.5]$-$[8.0]  
and [24] vs. [8.0]$-$[24] color-magnitude diagrams with boundaries indicating 
the two SWIRE-based sets of color-magnitude criteria.  Sources located in the 
upper right corner in {\it either} color-magnitude diagram are selected as 
YSO candidates, while those to the left are likely stars and those remaining 
are likely galaxies. (Note that the YSO candidate under the dashed lines in the 
extragalactic region in one color-magnitude diagram must be located above the 
dashed line in the other color-magnitude diagram.) 21 objects are identified
as YSOs based on these two color-magnitude diagrams. As in Figure 10, sources 
identified with plus symbols in Figure 11 are those around the L1251B core.  
After removing the likely extragalactic sources in Figure 10, the number of 
Class 0/I 
candidates is reduced to 4 and all of them are located around the L1251B core. 
Also, the number of Class II candidates reduces to 14 in the L1251E region and 
to 5 around the L1251B core.  
Three of the YSOs identified by the color-magnitude diagrams (Figure 11) do 
not fall into the criteria of Class 0/I or Class II in the color-color
diagram (Figure 10) and located far from L1251B. 
Figure 12 shows the spatial 
distribution of YSO candidates in L1251E, and the fluxes of these YSO 
candidates are listed in Table 2 and 3. The flux uncertainty 
is $\sim$15\%. 

In summary, we are adopting the color criteria as given in the c2d catalog
and their source classification. First, the c2d color-magnitude diagrams 
(Figure 11), which do not provide the information on Classes, are used to 
extract YSO candidates, and as a next step to classify the evolutionary stages
of the YSO candidates, the color-color diagram seen in Figure 10 is used.
Finally, 4 Class 0/I and 14 Class II candidates are classified in L1251E 
even though 21 YSO candidates are extracted from the c2d color-magnitude 
diagrams. 
The galaxy contamination rate of our YSO candidate selection criteria
obtained from the SWIRE data is 15 sources per square degree, thus
the 0.4\deg$\times$0.35\deg\ Spitzer image of L1251E should have 2.1 
mis-identified YSO candidates.  Therefore, excluding 3 YSO candidates
(including one known HH object) is consistent with the prediction of
our adopted criteria.

Among the sources around the L1251B core, Figures 10 and 11 indicate
that IRS1, IRS2, and source 16 are in Class 0/I, IRS3, IRS5, IRS6, and
sources 8 and 9 are in Class II, and sources 7, 10, 12, 13, and 14 are
stars.  We note that Source 11 is detected only in the IRAC 3.6 $\mu$m
and 4.5 $\mu$m bands.  In Figure 10, some sources with
[5.8]$-$[8.0] $>$ 1.5 have [3.6]$-$[4.5] $<$ 1; from the c2d criteria, all
of these, except for one source (see below), are likely background
galaxies.  We note that IRS1 has a much redder [3.6]$-$[4.5] color than
a [5.8]$-$[8.0] color, perhaps caused by absorption within its massive
envelope by the 10 $\mu$m silicate feature that overlaps with the IRAC
8.0 $\mu$m band (Allen et al. 2004).  Similar colors are seen in other
deeply embedded sources (Jorgensen et al. 2006; Dunham et al. 2006).

Color-color identification of evolutionary state is based on
presumptions about the physical characteristics of each object, and this
could lead to some misclassifications.  For example, the exceptional
source (noted above) in Figure 10 with [5.8]$-$[8.0] $>$ 1.5 and
[3.6]$-$[4.5] $<$ 1 may be a Class 0/I YSO with an very low envelope
density (see Fig. 1 of Allen et al. 2004). In addition, the YSO
candidate with [3.6]$-$[4.5]$=$11 and [5.8]$-$[8.0]$=$$-$0.1 also might be
a Class 0/I object with a very low central luminosity, a low envelope
density, and a large centrifugal radius.  Finally, the YSO candidate
with colors similar to those of Class III objects or stars (i.e.,
Class III/Star) may be a Class II source with a low disk accretion rate
and a low inclination.  We expect the number of misclassifications to be
minimal, though calibration of color-color diagrams through modeling the
SEDs of specific sources therein would be welcome.

Figure 13 is a $K$ vs. $K-[24]$ color-magnitude diagram which is used to check
the possibility of confusion between our sources and background galaxies.  
All YSO candidates detected clearly at 24 \micron\ are bright or red enough 
to be separated clearly from background galaxies.  IRS1, IRS2, IRS4, and 
source 16, which are in Class 0/I and were detected separately by MIPS are 
very red.
IRS1, IRS2, and IRS4 are clearly the youngest most embedded members of L1251B, 
and they are 
possibly still forming out of the rapidly rotating envelope and through 
interaction with outflows (see Paper II).  IRS3, IRS5, and IRS6, however, 
were not clearly detected at 24 \micron\ even though they are classified as 
Class II and are close in projection to IRS1, IRS2, and IRS4. 
These may comprise an older population located at the outer boundary 
of the dense L1251B region.  

Figure 14 shows stellar model fits on top of SEDs 
of sources around L1251B.
According to the c2d stellar model fits, sources 7, 10, 12, 13, 14, and 15 are 
possible foreground or background stars reddened by some extinction.  Based 
on the description of the source type by catalogs (Evans et al. 2006; Lai et al.
in preparation), IRS1, IRS2, IRS4, and source 16 are potential Class 0/I 
objects and IRS3, IRS5, IRS6, and sources 8 and 9 are possibly
Class II, star/disk objects.  These results are consistent with results from 
the color-color diagram shown in Figure 10.

The MIPS 70 \micron\ image for L1251E is shown in Figure 15a.  Some of the 
YSO candidates show diffuse or point source-like emission at 70 \micron, 
though the emission from the inner L1251B region is confused by IRS1, IRS2, and 
IRS4.  To estimate the 70 $\mu$m fluxes of these unresolved sources,
70 $\mu$m fluxes were measured within a 24 $\mu$m intensity contour chosen 
to separate the three sources (see Figure 15b). Given the 
ratios of flux between sources enclosed by this contour, the total 70 $\mu$m 
flux, as measured by annular photometry, was divided among the three sources.
(In applying this procedure, we have assumed that the spatial extent of the 
70 $\mu$m sources matches that of the 24 $\mu$m resolved sources.)  
Note that the extent of emission from IRS1 will influence the flux measurements
of IRS2 and IRS4. 
To mitigate the influence of the extended emission from IRS1, therefore, the 
IRS1 flux has been fitted with a Gaussian profile, and the relative fluxes from
IRS2 and IRS4 above the wings of the IRS1 profile have been measured.

We also applied the 70 $\micron$ flux ratios to estimate the 350, 450, and 
850 $\micron$ fluxes of the three sources.  In Table 4, we list flux densities 
from 1.2 to 850 \micron\ of IRS1, IRS2 and IRS4 and in Table 5, we list 
bolometric luminosities ($L_{bol}$), bolometric temperatures ($T_{bol}$), and
$L_{smm}/L_{bol}$, where $L_{smm}$ is the luminosity at $\lambda > 350$ \micron,
for these sources based on their SEDs.  
According to the results, IRS1 is the dominant luminosity 
source in L1251B by a factor of $\sim$10.
Young et al. (2003) found that L1251B should be in Class 0 based on 
$T_{bol}$ (64 K) and $L_{smm}/L_{bol}$ (0.01), which
were calculated from single dish submillimeter continuum observations,
and assuming only one heating source (IRAS 22376+7455). 
However, as seen in Table 5, IRS1, IRS2, and IRS4 are classified to Class I 
objects based on $T_{bol}$ ($> 70$ K) while they are still categorized to 
Class 0 based on $L_{smm}/L_{bol}$ ($> 0.005$). 
The bolometric luminosities of IRS1 and IRS2 include submillimeter
emission from their respective envelopes, but these may also include
such emission from the prestellar objects between IRS1 and IRS2.
(Such prestellar objects do not have emission at shorter wavelengths
to contribute to bolometric luminosities.)  Given the proximity of
these objects to IRS1, their possible contributions likely affect
the bolometric luminosity estimate of IRS1 more than any other in
L1251B source.  Hence, the luminosity of IRS1 of 10 \lsun\ is likely
an upper limit. In contrast, if fluxes at submillimeter wavelengths
($\lambda \geqq 350$ \micron) are not considered, the lower limit of the
luminosity of IRS1 is 5 \lsun.
To quantify the amount of contribution of fluxes at submillimeter
wavelengths from each source, detail models are necessary.

\section{DISCUSSION} 

In L1251E, about 30 point sources were detected in the MIPS 24 \micron\ band, 
and $\sim$75\% of these were classified as YSO candidates (see Figure 11).  
Based on color-color classifications (see Figure 10), about 22\% and 78\% 
of the YSOs in L1251E are in Class 0/I and Class II respectively.  (A few
YSO candidates, however, have colors that do not fall into the Class 0/I or 
Class II classifications.)  Within L1251B, however, 6 YSOs were detected 
half of them are classified as Class 0/I candidates, and the other half 
are classified as Class II protostars. The whole 
field size of the MIPS observation is about 2.5 pc$^2$, so the number density 
of YSOs in L1251E is about 10 pc$^{-2}$.  The spatial number density of YSOs 
in L1251B exceeds that in the whole L1251E region by two orders 
of magnitude.  
The area used for the number density calculation in L1251B is
$\sim$0.01 pc$^{2}$, which covers emission at submillimeter wavelengths
shown in Figure 6.

We compare L1251E and L1251B with two larger cluster forming regions
observed with the SST: NGC 2264 (Teixeira et al. 2006) and Perseus
(Jorgensen et al. 2006).  For this, we concentrate on the ``Spokes"
cluster in NGC 2264 and NGC 1333 and IC 348 in Perseus.
The Spokes cluster ($\sim$2 pc$^{2}$) and NGC 1333 ($\sim$3.2 pc$^{2}$)
have similar spatial areas to L1251E ($\sim$2.5 pc$^{2}$), and
IC 348 has about three times bigger area ($\sim$6.9 pc$^{2}$) than L1251E.

Forty YSOs were detected in the Spokes cluster, which is the densest 
part of NGC 2264, in the MIPS 24 \micron\ band.  
The area of the Spokes cluster is $\sim$2 pc$^{2}$, yielding a YSO 
number density of $\sim$20 pc$^{-2}$, greater than that of L1251E by a 
factor of 2.  
In NGC 2264, however, the magnitude of the weakest YSO detected 
at 24 \micron\ is about 6.7 mag.  With a similar restriction (and keeping in 
mind the factor of $\sim$2 in distance between regions), the number of point 
sources detected in the MIPS 24 micron band with [24] $<$ 6.7 in L1251E is 10, 
and all are classified as YSO candidates (see Figure 11).  Based on this 
number, the spatial number density of YSO candidates in L1251E is $\sim$4 
pc$^{-2}$, i.e., smaller than that in the Spokes cluster by a factor of 5.  
Despite this difference in spatial number densities, the fraction of Class 0/I 
candidates in L1251E ($\sim$22\%) is not very different.
In addition, the fraction of Class 0/I candidates in L1251B
(50\%), the densest region in L1251E, is not different from that of the dense 
region around IRAS 12 in NGC 2264 ($\sim$60\%).  

IC 348 and NGC 1333 show similar number densities of YSOs to that in L1251E. 
The fractions of Class 0/I candidates in IC 348 and NGC 1333 are about 14\% and 
36\%, respectively. In the remaining cloud outside IC 348 and NGC 1333, 
however, the fraction is about 47\%, indicative of different formation 
timescales over Perseus.
In summary, therefore, the number density of YSOs in L1251E is similar to those 
in active star forming regions of NGC 2264 and Perseus. 

L1251B has three Class 0/I members (IRS1, IRS2, and IRS4), which have been 
detected at 24 \micron, and three more Class II members (IRS3, IRS5, and IRS6) 
are located in projection inside L1251B in the IRAC images (e.g.,
see Figures 3 or 4).  
The average projected distance among 18 YSOs in L1251E is about 213\arcsec\
with the standard deviation of $\sim$148\arcsec\, whereas the average 
projected distance of 6 YSOs in L1251B is about 24\arcsec\, indicating 
clustering. 
Single dish submillimeter continuum emission, which traces dense envelope 
material, covers the 6 YSOs of L1251B with intensity peaks between IRS1
and IRS2. In addition, two sources located between IRS1 and IRS2 have been 
newly detected at 1.3 mm with the SMA. They are not associated to any IRAC or
MIPS source; and are possibly prestellar condensations. The shift of
intensity peaks in the submillimeter continuum emission (see Figure 6 and \S 3)
observed with single dish telescopes seems related to the variation of physical
properties of the two prestellar condensations. For example, the shift of 
intensity peaks from the north to the south with wavelength can be explained if 
the prestellar source at the north has a higher temperature than the other one.
These observed features covering various evolutionary stages (from prestellar 
to Class II) suggest that fragmentations during gravitational
collapse of a dense molecular core (Boss 1997; Machida et al. 2005) can 
explain the formation of this small group of pre- and protostellar objects
in L1251B. 
Class 0/I objects in L1251B might help the formation of further
prestellar condensations since the external pressure resulting from
a previous star formation event could reduce Jeans fragmentation
length scales in the remaining matter for the next incidence of
star formation (Ward-Thompson et al. 2006 and references).
We examine the molecular line data in Paper II
to understand how dynamics and chemistry are distributed in L1251B, and
how the group members of L1251B are interacting.

Many YSO candidates in various evolutionary stages are distributed all over 
the extended L1251E region.  In addition, the east core has a radius of about 
0.3 pc and may harbor future star formation (see Paper II).  
This suggests that L1251E is a possible example of low-mass cluster formation,
and L1251B, which is a small group of pre- and protostellar objects, is just 
part of the larger, on-going cluster formation.

\section{SUMMARY}

L1251E, the densest C$^{18}$O core of L1251, has been observed by the SST as 
part of the SST c2d Legacy project.  The most interesting region in L1251E is 
L1251B, which has been revealed by the SST to be a small group of protostars.
We have compared the SST data with other continuum data to 
study L1251B more comprehensively.  The summary of our results is as follows:

1. L1251E contains at least two cores, a western one coincident with L1251B 
and an eastern one detected at 850 \micron\ continuum.
No source has been detected in the IRAC or MIPS bands toward the east 
core, suggesting it is starless.  Asymmetrically blue line profiles (see Paper 
II) have been detected toward the eastern core, however, that are suggestive of 
infall motions.  

2. About 20 YSO candidates (4 in Class 0/I, 14 in Class II, and 3 unclassified 
YSOs) are distributed all over L1251E, suggesting L1251E is a possible 
example of low mass clustered star formation.

3. L1251E has a surface number density of young stellar objects of $\sim$10 
pc$^{-2}$, not very different from those seen in the Spokes cluster of 
NGC 2264 and the most active star forming regions of Perseus, IC 348 and 
NGC 1333. The spatial number density in L1251B is, however, larger than that of
L1251E by about two orders of magnitude, indicative of highly concentrated 
star formation.
The fractions of Class 0/I candidates in aL1251E and L1251B are similar to
those in and around the Spokes cluster, but the fraction in Perseus varies 
from region to region.

4. Six sources located in projection within L1251B were detected in all 
IRAC bands but only IRS1, IRS2 and IRS4 are detected clearly at 24 \micron\ 
(with MIPS) and classified as Class 0/I candidates.
IRS2 is associated with a near-infrared bipolar nebula. 
   
5. Dust continuum emission maps made at 350 \micron\ and 450 \micron\ show 
two intensity peaks in L1251B.  The stronger peak is not associated with any 
IRAC source but the weaker peak is associated with IRS1, the brightest 
object in L1251B.  The stronger peak is located between IRS1 and IRS2.   
The weaker peak disappears at 850 \micron\ and 1300 \micron\ and the 
distribution of continuum emission around the stronger peak becomes more 
spherical (and shifts to the south with increasing wavelength.)  Therefore, 
the stronger peak seems to be a column density peak, and the weaker peak is 
possibly from dust heated by IRS1.

6. IRS1 and IRS2 are detected as compact objects in the OVRO 3 mm and the SMA
1.3 mm observations, indicating that they are deeply embedded disk sources.
In addition, two sources have been newly detected between IRS1 and IRS2 at
1.3 mm with the SMA. These two sources confirm that the stronger intensity 
peaks at 350 and 450 \micron\ and intensity peaks at 850 and 1300 \micron\ are 
column density peaks. 
The positional shift of continuum density peaks with wavelength is possibly 
caused by a temperature difference between the two newly detected sources.

\acknowledgments
Support for this work, part of the {\it Spitzer} Legacy Science Program, 
was provided by NASA through contracts 1224608 and 1230782 issued by the Jet 
Propulsion Laboratory, California Institute of Technology, under NASA contract 
1407.  This work was also supported by NASA Origins grant NNG04GG24G.  Support 
for this work was also provided by NASA through Hubble Fellowship grant 
HST-HF-01187 awarded by the Space Telescope Science Institute, which is 
operated by the Association of Universities for Research in Astronomy, Inc.,
for NASA, under contract NAS 5-26555.  We are very grateful to Mario Tafalla 
for providing the unpublished 1.3 mm continuum map. 
We also thank Robert Gutermuth for supplying the IDL code used to 
make Figure 4.  Jeong-Eun Lee thanks the University of Texas at Austin 
for the support through the University Continuing Fellowship.
We are very grateful to Geoff Blake, Lee Mundy, and the referee of this paper, 
Paul Ho for many helpful comments.


\clearpage

\begin{deluxetable}{cccccc}
\rotate
\tablecolumns{8}
\tablewidth{0pc}
\tablecaption{OVRO and SMA Observational Summary}
\tablehead{
\colhead{} & \colhead{} & \colhead{Gaussian} &
\colhead{Synthesized} & \colhead{Synthesized} \\
\colhead{} & \colhead{Bandwidth} & 
\colhead{Taper FWHM} & \colhead{Beam FWHM} & \colhead{Beam P.A.} & 
\colhead{1 $\sigma$ rms\tablenotemark{a}} \\
\colhead{Tracer} & \colhead{(GHz)} &  
\colhead{(\as\ $\times$ \as)} & 
\colhead{(\as\ $\times$ \as)} & \colhead{(\deg)} & \colhead{(Jy beam$^{-1}$)}}
\startdata
OVRO 1.33 mm    &  2\tablenotemark{b}  & 
1.60 $\times$ 1.60  & 2.4 $\times$ 2.3   &  309  &  0.003  \\
OVRO 2.95 mm    &  2\tablenotemark{b}  & 
3.25 $\times$ 3.25  & 5.2 $\times$ 4.8   &  321  &  0.0007 \\
\cr
SMA 1.33 mm    &  4\tablenotemark{b}  & 
\nodata  & 4.1 $\times$ 3.5  &  339  &  0.003 \\
\enddata
\tablenotetext{a}{1 $\sigma$ rms computed from noise-free regions of the 
deconvolved maps.}
\tablenotetext{b}{Continuum band widths of the OVRO and SMA consist of 1 and 
2 GHz, respectively, each from the LSB and USB.}
\end{deluxetable}

\begin{deluxetable}{lccccccc}
\rotate
\tablecolumns{9}
\tabletypesize{\footnotesize}
\tablecaption{\bf Class 0/I candidates in L1251E  
\label{tab3}}
\tablewidth{0pt}
\tablehead{
\colhead{RA (J2000)}    &
\colhead{Dec (J2000)}              &
\colhead{}              &
\colhead{}              &
\colhead{Fluxes (mJy)}              &
\colhead{}              &
\colhead{}       &
\colhead{ID\tablenotemark{a}}  \\     
\colhead{($^h$ $^m$ $^s$)}    &
\colhead{($\degree$ $\arcmin$ $\arcsec$)}              &
\colhead{IRAC 1}              &
\colhead{IRAC 2}              &
\colhead{IRAC 3}              &
\colhead{IRAC 4}              &
\colhead{MIPS 1}    &   
\colhead{}       
}
\startdata
 22 38 42.8 & 75 11 36.8 & 6.54e+00& 9.06e+00& 7.02e+00& 1.28e+01& 3.95e+02 & 
L1251B-IRS4 \\
 22 38 46.9 & 75 11 33.9 & 9.37e+00& 1.25e+02& 2.74e+02& 3.63e+02& 3.12e+03 & 
L1251B-IRS1 \\
 22 38 53.0 & 75 11 23.5 & 5.76e+00& 1.09e+01& 1.27e+01& 1.45e+01& 4.74e+02 & 
L1251B-IRS2 \\
 22 39 13.3 & 75 12 15.8 & 4.01e+01& 1.13e+02& 2.01e+02& 2.10e+02& 3.48e+02 & 16 \\
\enddata
\tablenotetext{a}{Source ID marked in Figure 9.
}
\end{deluxetable}

\begin{deluxetable}{lccccccc}
\rotate
\tablecolumns{9}
\tabletypesize{\footnotesize}
\tablecaption{\bf Class II candidates in L1251E
\label{tab4}}
\tablewidth{0pt}
\tablehead{
\colhead{RA (J2000)}    &
\colhead{Dec (J2000)}              &
\colhead{}              &
\colhead{}              &
\colhead{Fluxes (mJy)}              &
\colhead{}              &
\colhead{}       &
\colhead{ID\tablenotemark{a}}  \\
\colhead{($^h$ $^m$ $^s$)}    &
\colhead{($\degree$ $\arcmin$ $\arcsec$)}              &
\colhead{IRAC 1}              &
\colhead{IRAC 2}              &
\colhead{IRAC 3}              &
\colhead{IRAC 4}              &
\colhead{MIPS 1}    &
\colhead{}
}
\startdata
22 37 49.6 & 75 ~4 ~6.4 & 1.85e+01& 1.30e+01& 9.02e+00& 7.04e+00& 9.55e+01\tablenotemark{c}& \\
22 38 11.6 & 75 12 14.6 & 8.45e+00& 8.13e+00& 7.83e+00& 8.83e+00& 2.88e+01& \\
22 38 15.2 & 75 ~7 20.4 & 1.39e+01& 1.32e+01& 9.82e+00& 9.20e+00& 1.73e+01& \\
22 38 18.7 & 75 11 53.8 & 1.22e+02& 1.63e+02& 2.32e+02& 2.50e+02& 3.22e+02& \\
22 38 29.6 & 75 14 26.7 & 8.45e+00& 7.16e+00& 6.32e+00& 7.45e+00& 1.09e+01& \\
22 38 40.5 & 75 ~8 41.3 & 7.34e+00& 6.83e+00& 5.84e+00& 5.87e+00& 9.64e+00& \\
22 38 42.5 & 75 11 45.5 & 1.62e+01& 1.32e+01& 1.17e+01& 1.05e+01& 
5.66e+01\tablenotemark{b}& L1251B-IRS3\\
22 38 44.0 & 75 11 26.8 & 9.91e+00& 8.20e+00& 5.46e+00& 4.73e+00& 
3.91e+01\tablenotemark{b}& L1251B-IRS5\\
22 38 48.1 & 75 11 48.8 & 5.68e+00& 6.22e+00& 6.54e+00& 6.02e+00&
-1.92E+00\tablenotemark{b}& L1251B-IRS6\\
22 38 50.7 & 75 10 35.3 & 4.10e+00& 3.53e+00& 3.03e+00& 2.53e+00& 3.97e+00& 8 \\
22 39  4.7 & 75 11 ~1.2 & 1.71e+00& 1.64e+00& 1.60e+00& 1.79e+00& 3.57e+00& 9 \\
22 39 27.2 & 75 10 28.4 & 3.74e+01& 3.24e+01& 2.88e+01& 3.04e+01& 6.99e+01\tablenotemark{d}& \\
22 39 40.3 & 75 13 21.6 & 1.82e+01& 1.80e+01& 1.86e+01& 1.66e+01& 1.11e+01\tablenotemark{e}& \\
22 39 46.4 & 75 12 58.7 & 2.15e+02& 1.66e+02& 2.43e+02& 2.43e+02& 2.84e+02& \\
\enddata
\tablenotetext{a}{Source ID marked in Figure 9.
}
\tablenotetext{b}{The upper limit of the MIPS 24 \micron\ flux. 
}
\tablenotetext{c}{IRAS 22367+7488 (Kun \& Prusti 1993) 
}
\tablenotetext{d}{T Tau-type star (Kun \& Prusti 1993) 
}
\tablenotetext{e}{IRAS 22385+7456 (Kun \& Prusti 1993) 
}
\end{deluxetable}

\begin{deluxetable}{lccc}
\tablecolumns{4}
\footnotesize
\tablecaption{\bf Flux densities (mJy) of L1251B sources
\label{tab5}}
\tablewidth{0pt}
\tablehead{
\colhead{Wavelength (\micron)}                &
\colhead{L1251B-IRS1}    &
\colhead{L1251B-IRS2}              &
\colhead{L1251B-IRS4}              
}
\startdata
1.235 &  9.47e-02 & 5.55e-01 & 2.13e-01\\
1.662 &  1.03e+00 & 1.89e+00 & 6.04e-01\\
2.159 &  3.52e+00 & 4.25e+00 & 1.83e+00\\
3.555 &  9.37e+00 & 5.76e+00 & 6.54e+00\\
4.493 &  1.25e+02 & 1.09e+01 & 9.06e+00\\
5.731 &  2.74e+02 & 1.27e+01 & 7.02e+00\\
7.872 &  3.63e+02 & 1.45e+01 & 1.28e+01\\
24.00 &  3.12e+03 & 4.74e+02 & 3.95e+02\\
70.00\tablenotemark{a} & 2.5e+04 & 8.5e+02 & 1.4e+03 \\
350.0\tablenotemark{a} & 4.2e+04 & 2.8e+03 & 4.0e+03 \\
450.0\tablenotemark{a} & 2.1e+04 & 1.3e+03 & 2.1e+03 \\
850.0\tablenotemark{a} & 5.8e+03 & 3.4e+02 & 5.8e+02 \\
\enddata
\tablenotetext{a}{Aperture sizes for total 70, 350, 450, and
850 \micron\ fluxes covering three sources were 50\as, 60\as, 120\as, and 
120\as respectively.}
\end{deluxetable}

\begin{deluxetable}{lccc}
\tablecolumns{4}
\footnotesize
\tablecaption{\bf Luminosities and bolometric temperatures of L1251B sources
\label{tab6}}
\tablewidth{0pt}
\tablehead{
\colhead{}                &
\colhead{L1251B-IRS1}    &
\colhead{L1251B-IRS2}              &
\colhead{L1251B-IRS4}              
}
\startdata
$\rm L_{bol}$ (\lsun) &  10  &  0.6  & 0.8  \\
$\rm T_{bol}$ (K) & 87  & 140  & 98  \\
$\rm L_{smm}\tablenotemark{a}/L_{bol}$ & 0.03  & 0.03  & 0.04  \\
\enddata
\tablenotetext{a}{luminosity at $\lambda > 350$ \micron}
\end{deluxetable}
\clearpage

\begin{figure}
\figurenum{1}
\plotone{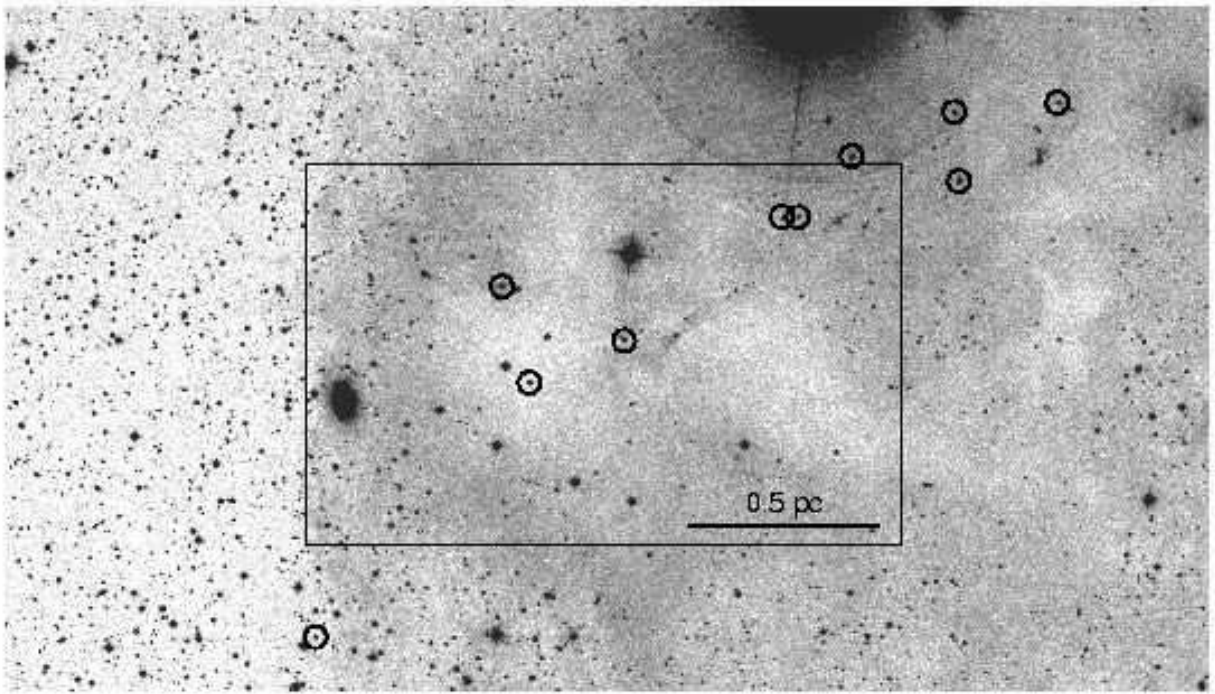}
\caption{
The Digitized Sky Survey (DSS) $R-$band image of L1251E.  The box indicates 
the field of view of Figure 2 and the circles indicate H$\alpha$ stars from 
Kun \& Prusti (1993).
}
\end{figure}

\begin{figure}[t]
\centering
\figurenum{2}
\includegraphics[width=6in]{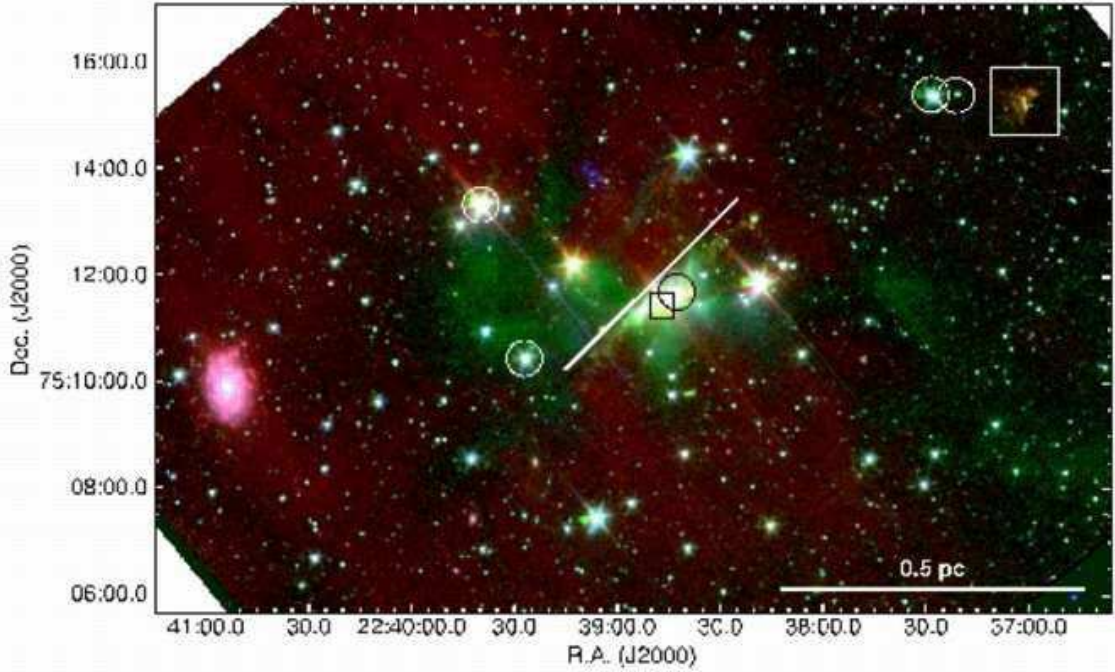}
\caption{
Three-color composite SST image of L1251E.
The IRAC 3.6, 4.5,
and 8.0 $\micron$ data are presented respectively as blue, green, and red.
The small box represents the position of IRAS 22376+7455 and the large box
includes the Herbig-Haro (HH) object, HH 373. The circles denote the same 
H$\alpha$ stars shown in Figure 1.  The H$\alpha$ star with the black box 
seems consistent with IRS3 (Figure 9). The red spiral galaxy located at 
(22$^{h}$40$^{m}$54.6$^{s}$, +75\deg 09\am 52.5\as) is UGC 12160 (Cotton 
et al. 1999). The white line, which is in the direction of SE-NW and centered 
around IRSA 22376+7455, indicates a jet-like feature, $\sim$0.3 pc long.
}
\end{figure}

\figurenum{3}
\begin{figure}
\plotone{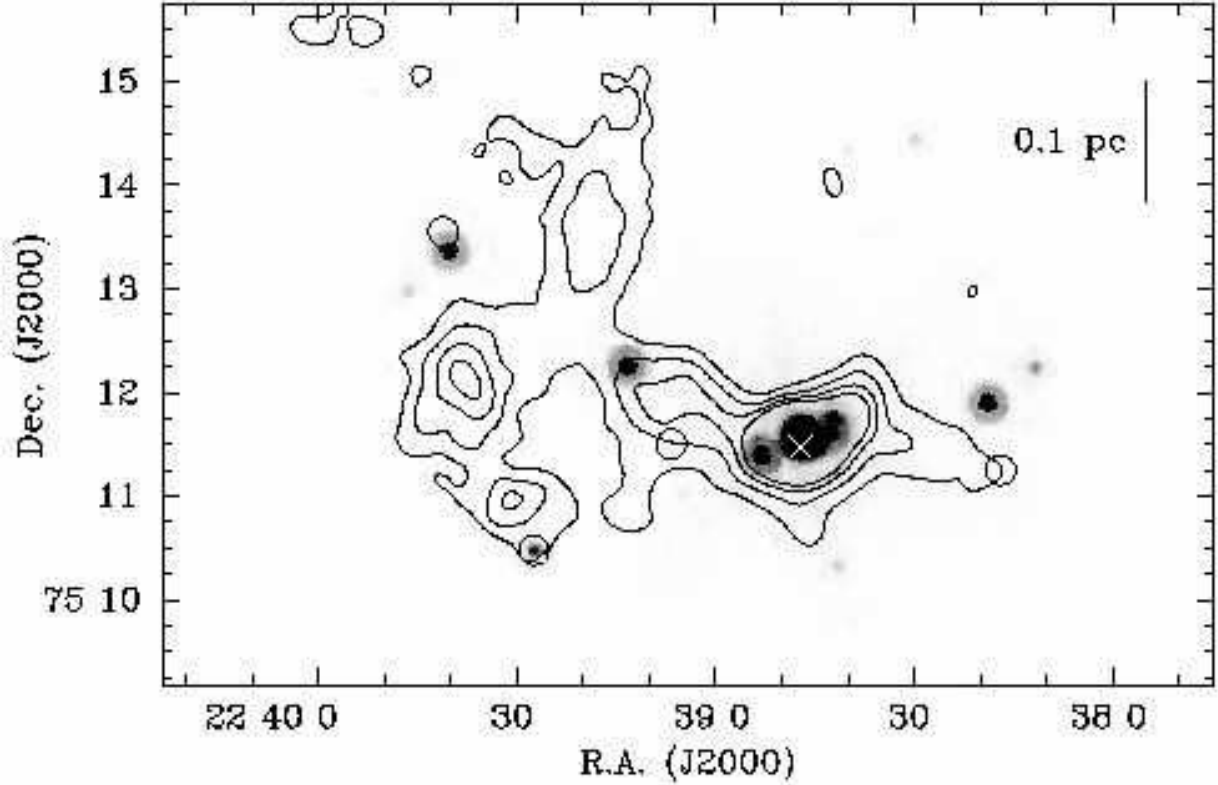}
\caption{
Mid-infrared and submillimeter continuum emission in L1251B.  Emission
at $\lambda$ = 24 $\mu$m obtained with MIPS is shown in greyscale while
emission at $\lambda$ = 850 $\mu$m obtained with SCUBA is shown in black
contours.  The greyscale range is 15-50 MJy sr$^{-1}$ and the contours
begin at 0.05 Jy beam$^{-1}$ and increase in steps of 0.05 Jy beam$^{-1}$.
To avoid confusion inside L1251B where emission is very strong and three
sources are detected at $\lambda$ = 24 $\mu$m, only 4 contours are used.
The white X denotes the position of IRAS 22376+7455 and the open circles
denote other IRAS sources found by Kun \& Prusti (1993).  
}
\end{figure}

\begin{figure}
\figurenum{4}
\epsscale{1}
\plotone{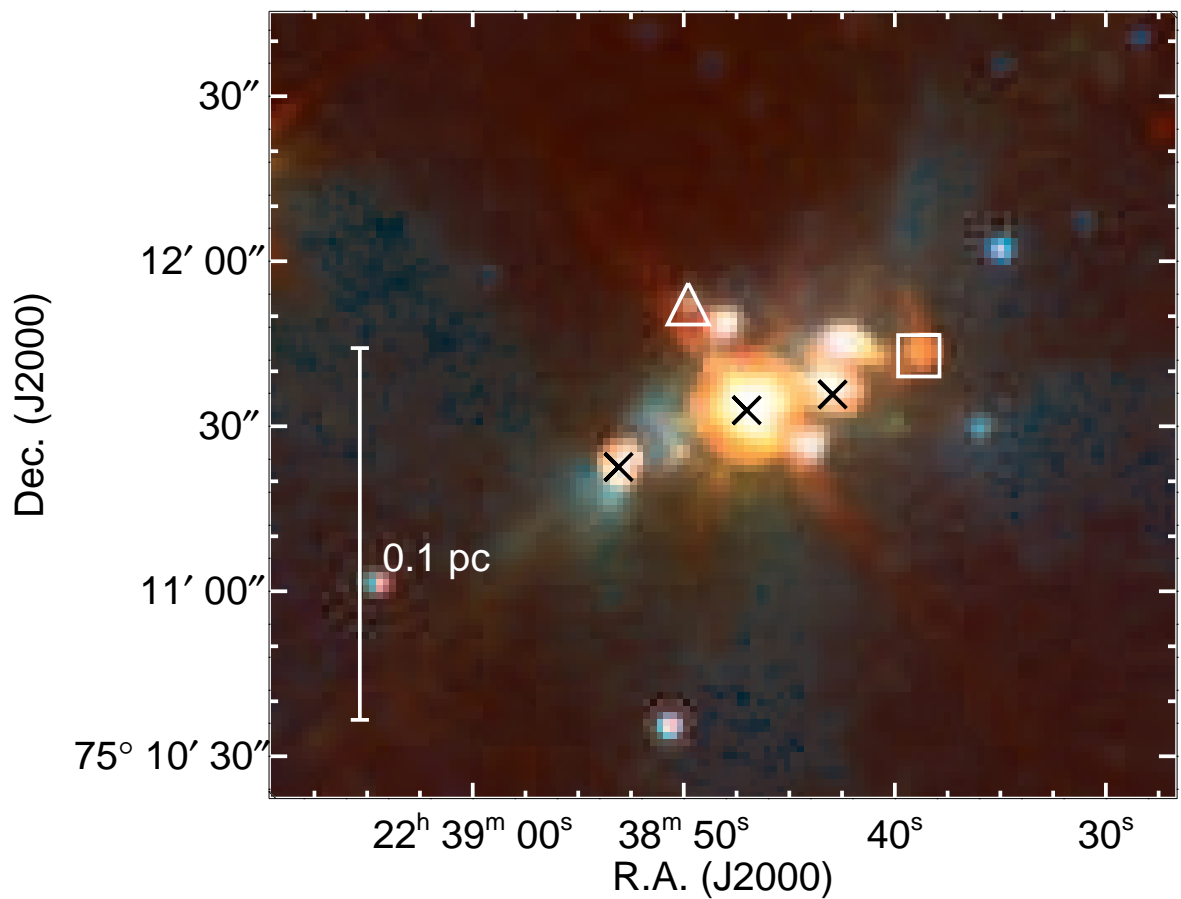}
\caption{
The zoomed-in three-color SST composite image around L1251B.  The color scheme 
is the same as in Figure 2.  The red source denoted with a square at 
(22$^{h}$38$^{m}$38.7$^{s}$, +75\deg 11\am 43.9\as) is weak and nebulous 
in the short IRAC bands (see Figure 5).  The source marked with a triangle 
is detected as a point source only at 3.6 \micron\ and 4.5 \micron. The 
X symbols indicate sources detected clearly at 24 \micron. }
\end{figure}
\clearpage

\begin{figure}
\figurenum{5}
\plotone{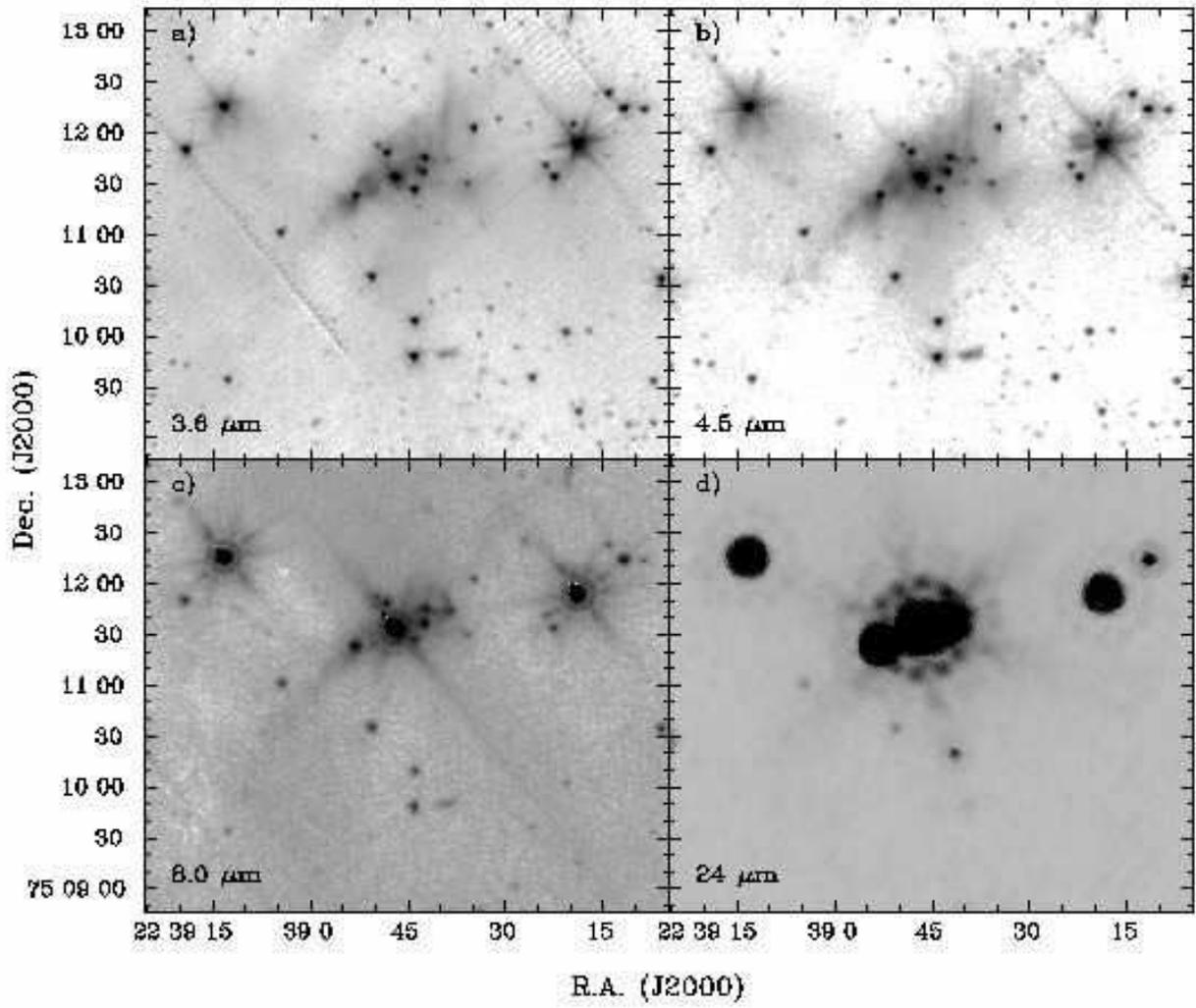}
\figcaption{
IRAC 3.6, 4.5, and 8.0 \micron\ and the MIPS 24 \micron\ images of L1251B.
}
\end{figure}

\begin{figure}
\figurenum{6}
\epsscale{1.00}
\plotone{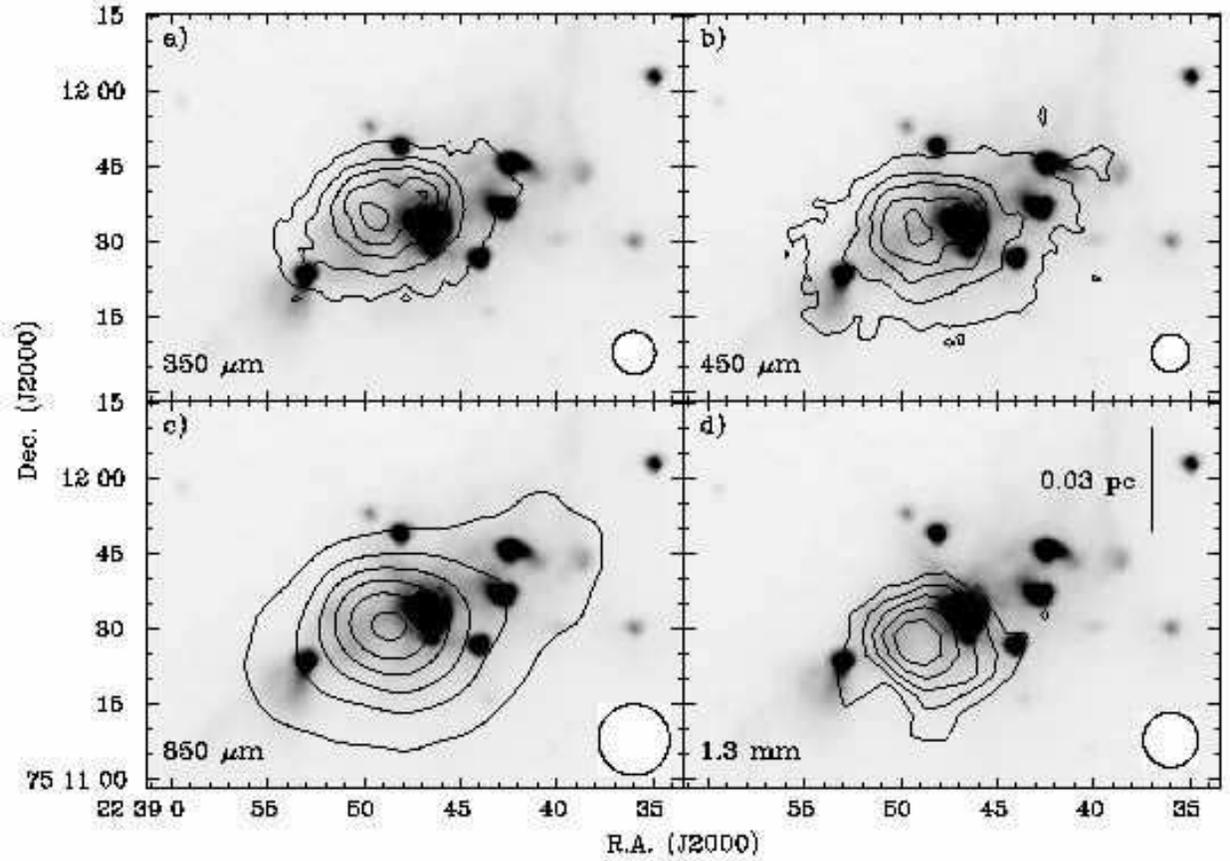}
\figcaption{
Comparisons of submillimeter continuum emission (contours) with near-infrared 
continuum emission (the IRAC 4.5 \micron\ band; greyscale) toward L1251B.  
The greyscale range is -3 to 5 MJy sr$^{-1}$.  a) Emission at $\lambda$ = 350 
$\mu$m obtained with SHARCII at the CSO.  Contour levels begin at 4 $\sigma$ 
and increase in steps of 5 $\sigma$ where 1 $\sigma$ = 0.04 Jy beam$^{-1}$.  
b) Emission at $\lambda$ = 450 $\mu$m obtained with SCUBA at the JCMT by Young 
et al. (2003).  Contour levels begin at 5 $\sigma$ and increase in steps of 
5 $\sigma$.  c) Emission at $\lambda$ = 850 $\mu$m obtained with SCUBA at the 
JCMT by Young et al. (2003).  Contour levels begin at 10 $\sigma$ and increase 
in steps of 10 $\sigma$.  d) Emission at $\lambda$ = 1.3 mm obtained with 
MAMBO at the IRAM 30 m Telescope.  Contour levels begin at 3 $\sigma$ and 
increase in steps of 3 $\sigma$ = 0.033 Jy beam$^{-1}$.
}
\end{figure}

\begin{figure}
\figurenum{7}
\epsscale{1}
\plotone{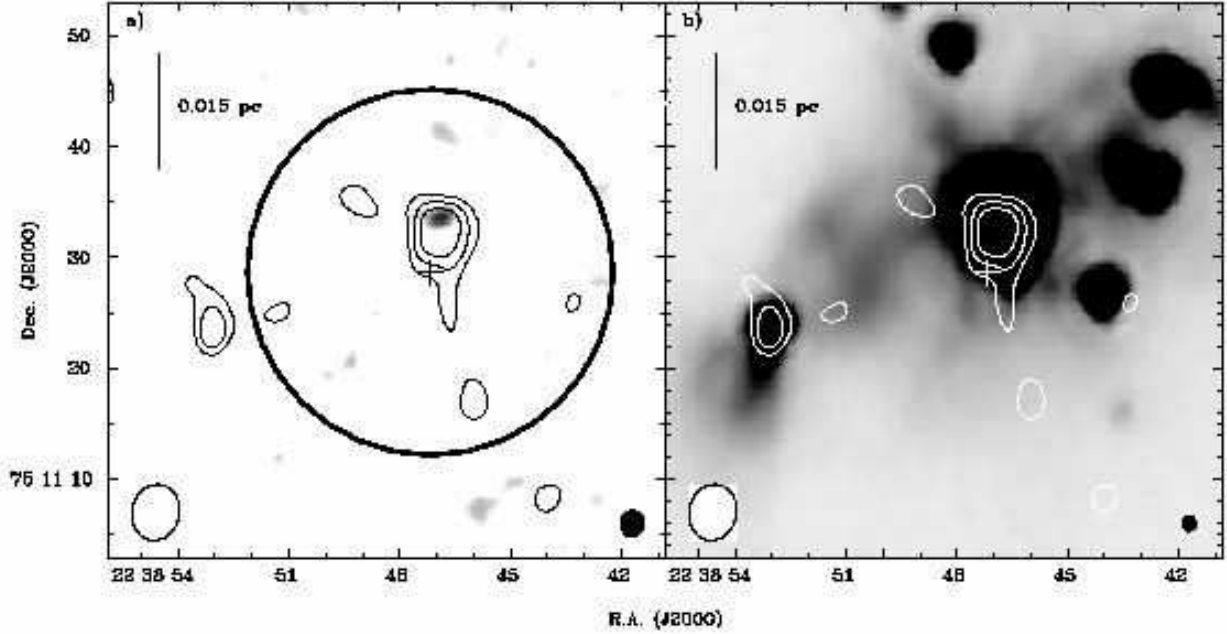}
\caption{
Interferometric maps of continuum emission toward L1251B, where the (0,0) position is defined
at the position of IRAS 22376+7455.
a) Emission
at $\lambda$ = 2.95 mm is shown as black contours while that at $\lambda$ = 
1.33 mm is shown in greyscale.  Contours begin at 2 $\sigma$ and increase 
in steps of 1 $\sigma$ = 0.0007 Jy beam$^{-1}$. The greyscale range is 
2-5 $\sigma$ where 1 $\sigma$ = 0.003 Jy beam$^{-1}$.  The cross denotes the
position of IRAS 22376+7455.  The large circle denotes the 32\farcs4 FWHM of 
the OVRO primary beam at $\lambda$ = 1.33 mm.  The ellipses at lower left and 
right denote relative sizes of the synthesized beams at $\lambda$ = 2.95 mm 
and $\lambda$ = 1.33 mm respectively.  b) Emission at $\lambda$ = 2.95 mm 
is shown as white contours as defined for panel a while that at IRAC Band 2 
($\lambda$ = 4.5 $\mu$m) is shown in greyscale.  The greyscale range is 
-0.17 -- 5.0 MJy sr$^{-1}$.  The cross and lower left ellipse are defined 
as in panel a, but the black ellipse at lower right denotes the resolution 
of the {\it Spitzer} Space Telescope at 4.5 $\mu$m.  Note that the secondary 
peak at $\lambda$ = 2.95 mm, seen $\sim$25\as\ SE of the central bright peak
(IRS1), is coincident with the IRAC source associated with the bipolar nebula 
(IRS2).
}
\end{figure}

\begin{figure}
\figurenum{8}
\epsscale{1.0}
\plotone{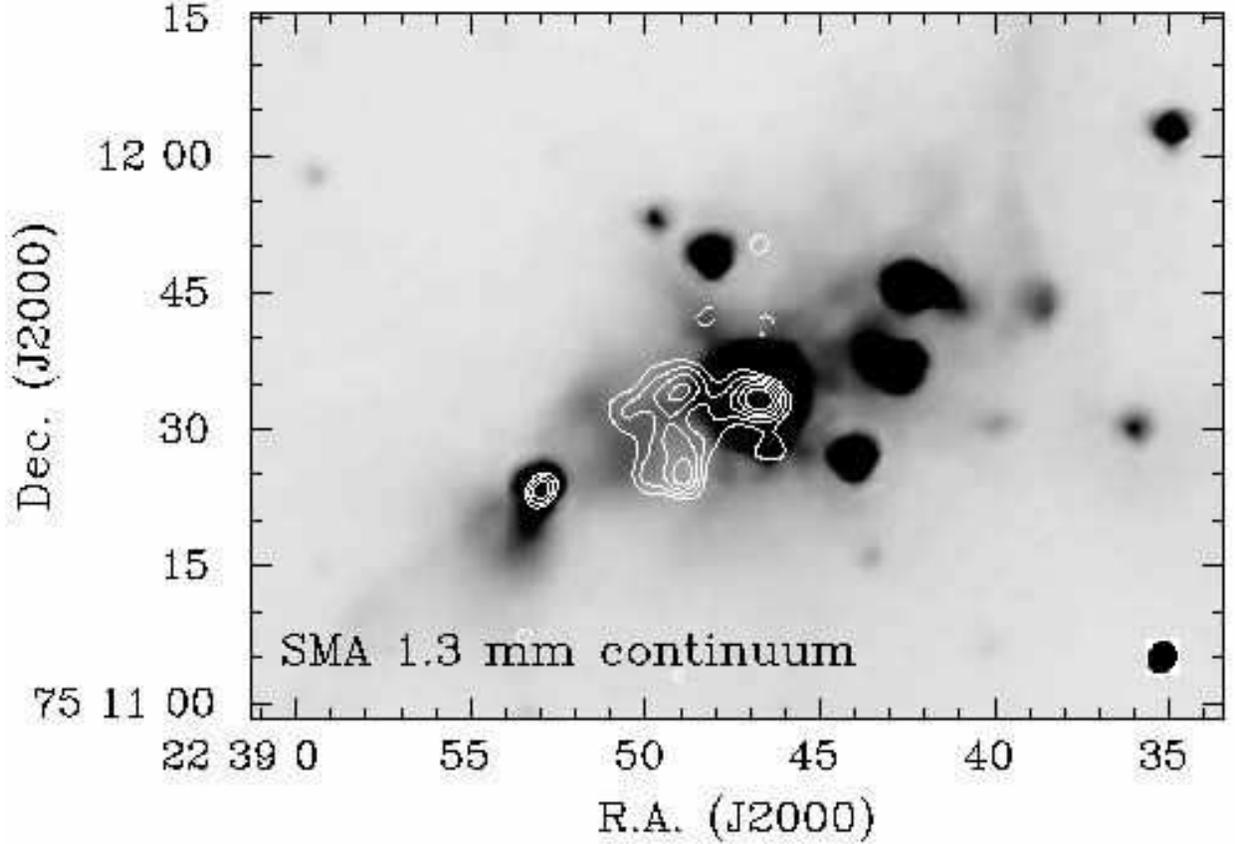}
\figcaption{
Comparison of the distribution of YSOs detected in the IRAC 4.5 \micron\ band
(gray scale) with the distribution of dust continuum emission (contours) 
observed at 1.3 mm with the SMA array. Contours begin at 12 mJy beam$^{-1}$
and increase in steps of 4 mJy beam$^{-1}$. Two new continuum peaks were
detected between IRS1 and IRS2. The locations of the new peaks are consistent
with the intensity peaks of N$_2$H$^+$ (see Paper II) and submillimeter
continuum emission observed with single dish telescopes.
}
\end{figure}

\begin{figure}
\figurenum{9}
\epsscale{1.0}
\plotone{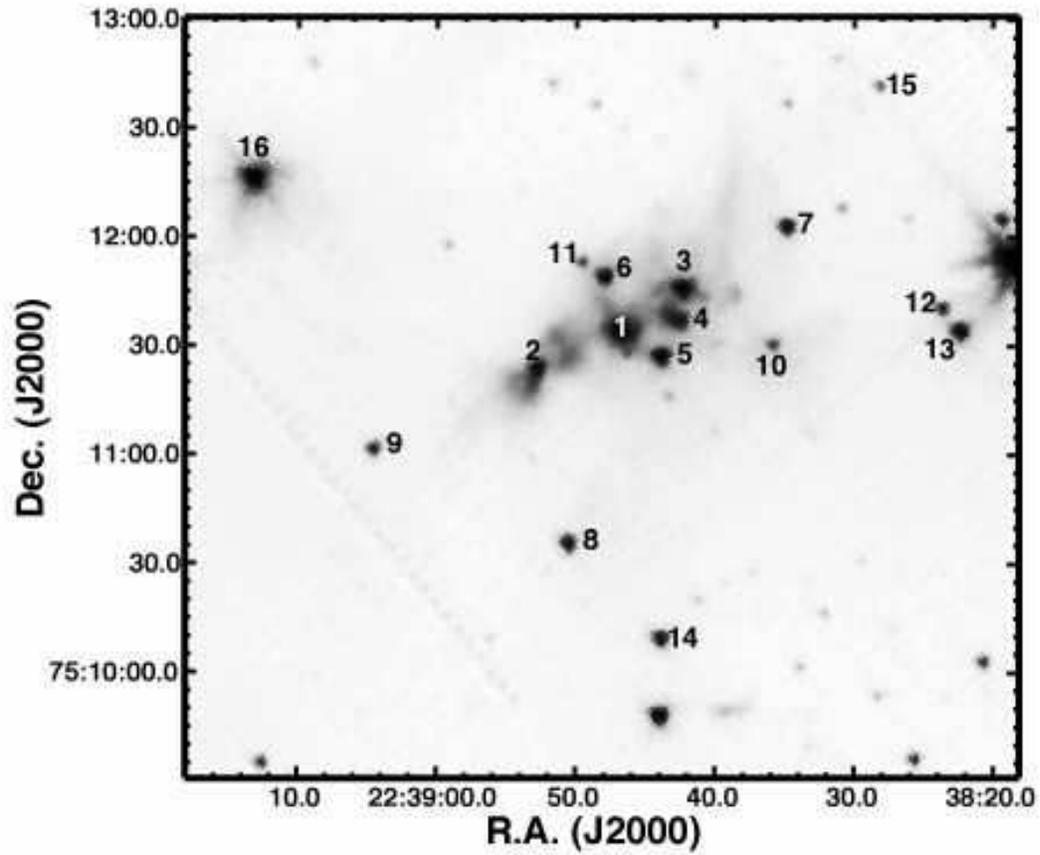}
\figcaption{
Source identification numbers around L1251B in the IRAC 3.6 \micron\ image.
For six sources (from 1 to 6) within L1251B, numbers are assigned along with 
R.A. after IRS1 and IRS2, which are the brightest source and the bipolar nebula 
source, respectively. 
Numbers (from 7 to 16) outside L1251B are, however, randomly assigned.
}
\end{figure}

\begin{figure}
\figurenum{10}
\plotone{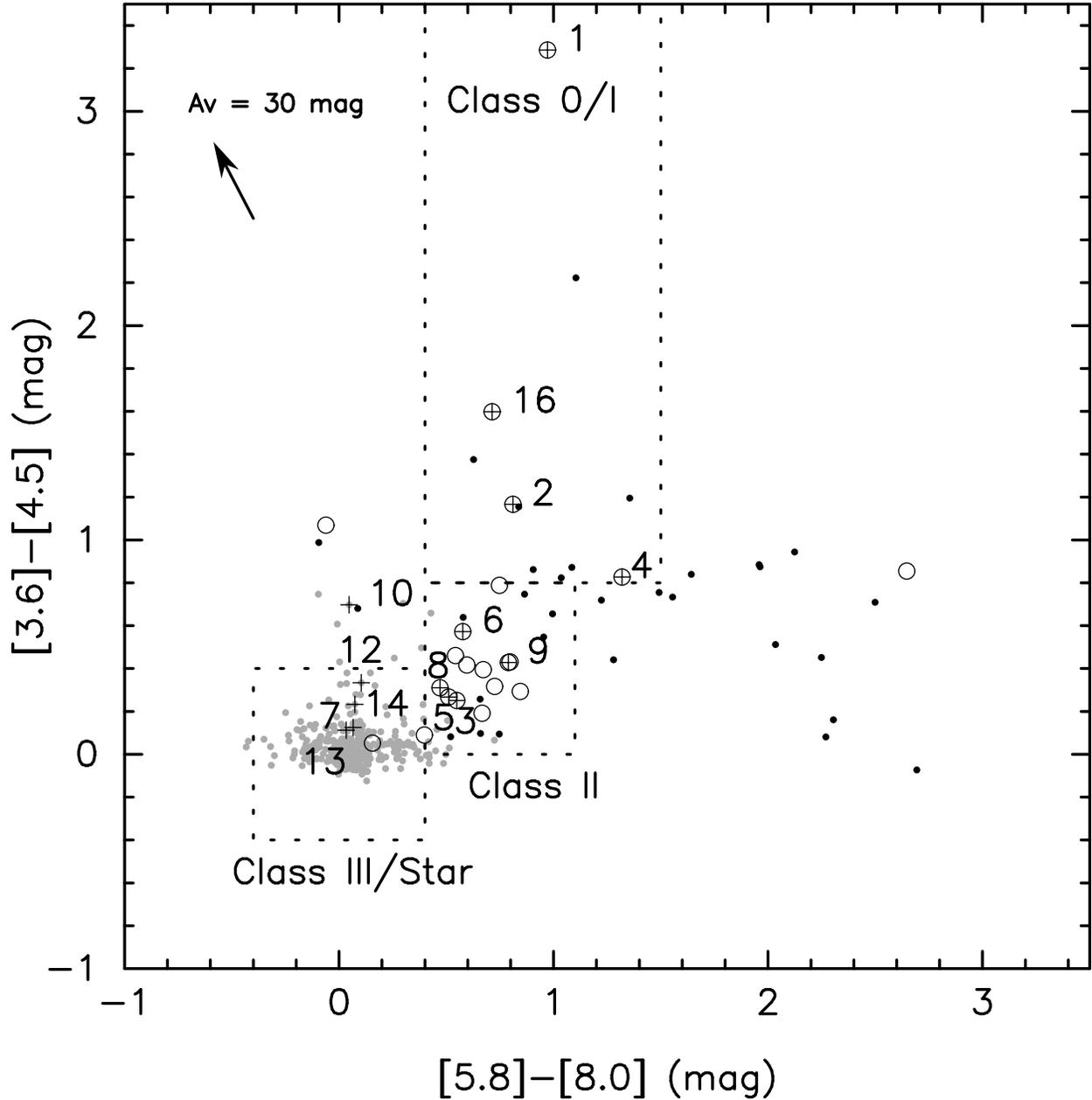}
\figcaption{
Color-color diagram for sources in the IRAC bands ($[3.6]-[4.5]$ vs. 
$[5.8]-[8.0]$).  The numbers associated with pluses indicate sources 
identified as in Figure 9.  Grey and black dots denote stars and 
galaxies, respectively, and open circles indicate YSO candidates, based 
on the c2d classification (see Figure 11).  The dashed boxes delineate 
the approximate domains of Class 0/I, Class II, and Class III sources 
based on Allen et al. (2004).  The open circles that are not included in 
Class 0/I and Class II are classified as YSO candidates mainly due to 
their MIPS 24 \micron\ fluxes, and the open circles, especially, with 
$[5.8]-[8.0] > 1.5$ are possibly extragalactic objects.  An extinction 
vector, which is calculated from Weingartner \& Draine (2001) model 
with $R_V=5.5$, is shown for $A_V=30$ mag.  
}
\end{figure}

\begin{figure}
\figurenum{11}
\plotone{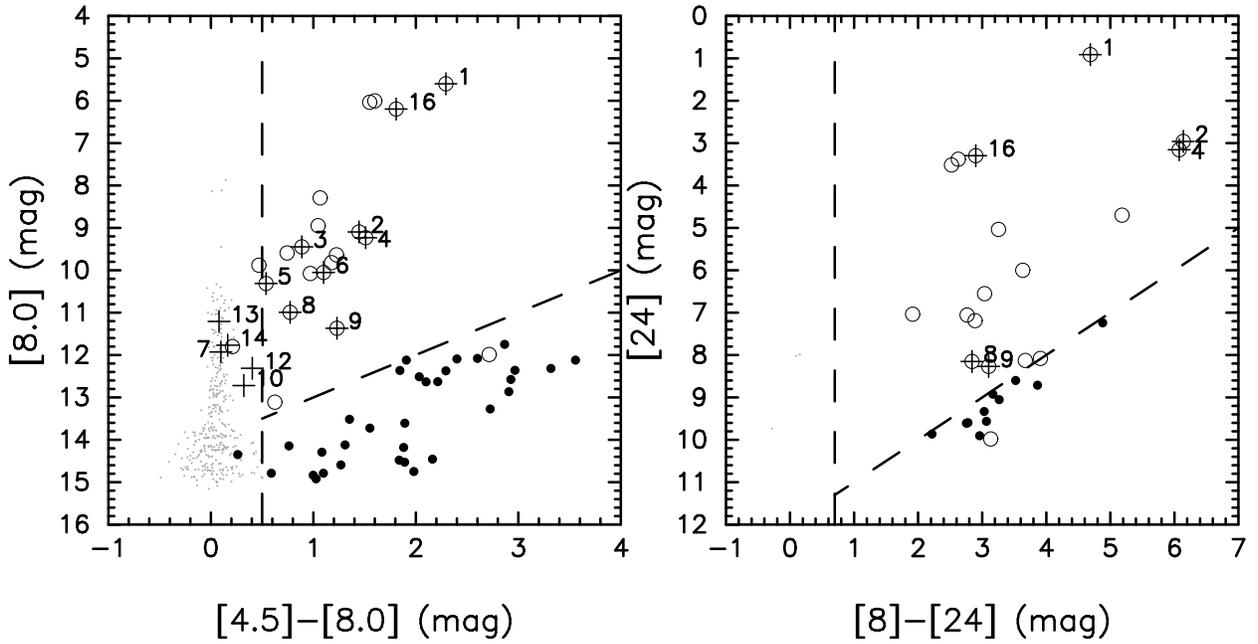}
\figcaption{
The c2d classification of detected sources based on two color-magnitude 
diagrams.  The left panel is an [8.0] vs. [4.5]$-$[8.0] color-magnitude 
diagram and the right panel is a [24] vs. [8.0]$-$[24] color-magnitude 
diagram.  The grey dots are ``stars" while open circles, located at the 
upper right part of either diagram, are YSO candidates (see \S 4).  
The black dots are everything else.
}
\end{figure}

\begin{figure}
\figurenum{12}
\plotone{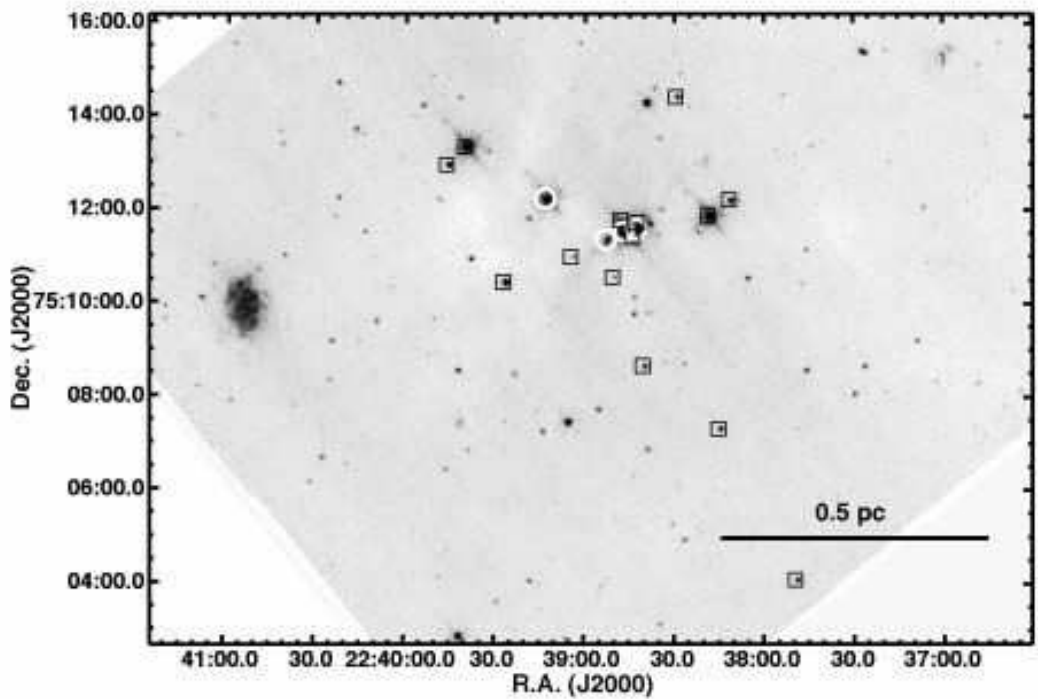}
\figcaption{
The spatial distribution of YSO candidates in L1251E, based on the c2d 
classification criteria. The grey map is the IRAC 8.0 \micron\ image, and
the open circles and boxes indicate Class 0/I and Class II candidates, 
respectively.
YSOs seem to follow a NE-SW chain, almost perpendicular to the jet direction
shown with the white line (NW-SE) in Figure 1.
}
\end{figure}

\begin{figure}
\figurenum{13}
\plotone{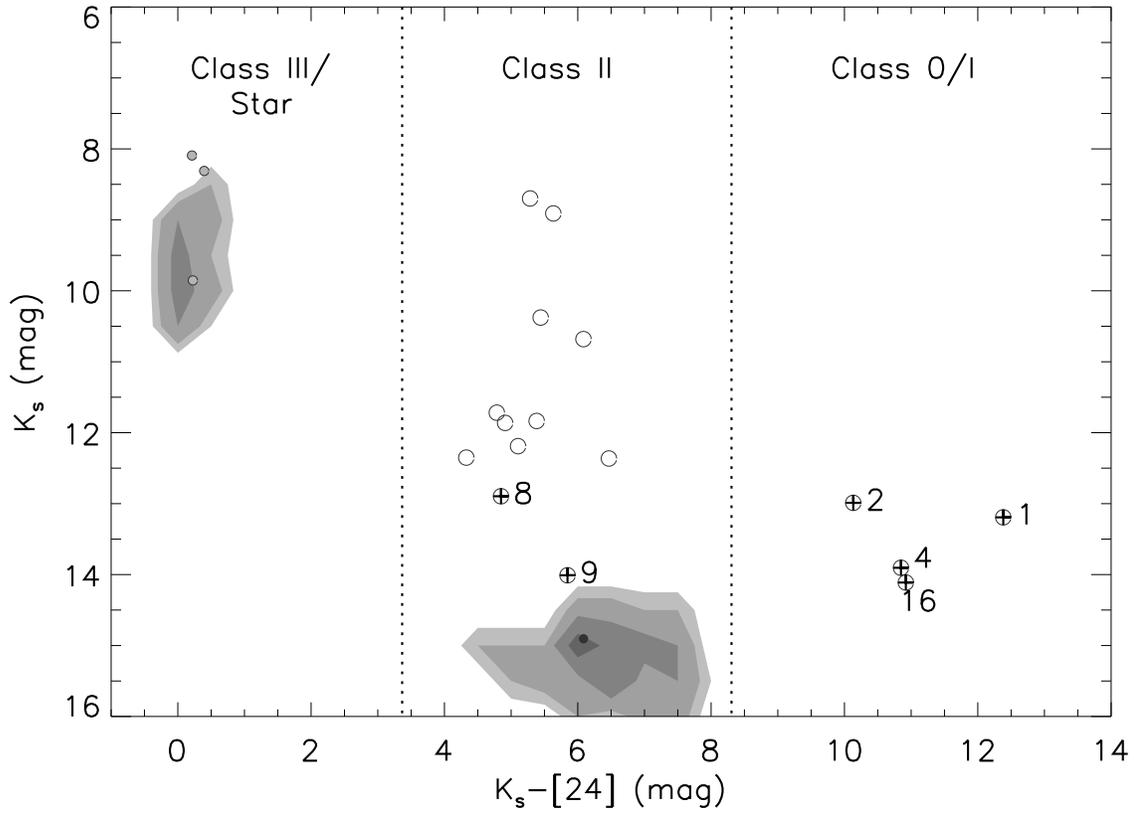}
\figcaption{
Color-magnitude diagram for the 2MASS $K$-band and the MIPS 24 $\micron$ band
sources. Grey contours present the number densities (1, 2, 4, and 7 per
degree$^2$) of background galaxies from the SWIRE data. The numbers with
pluses represent sources marked in Figure 9. The symbols are the same as 
in Figure 10 and 11 except that grey dots for stars have black boundaries to
distinguish them better.  
}
\end{figure}

\clearpage

\begin{figure}
\figurenum{14}
\plotone{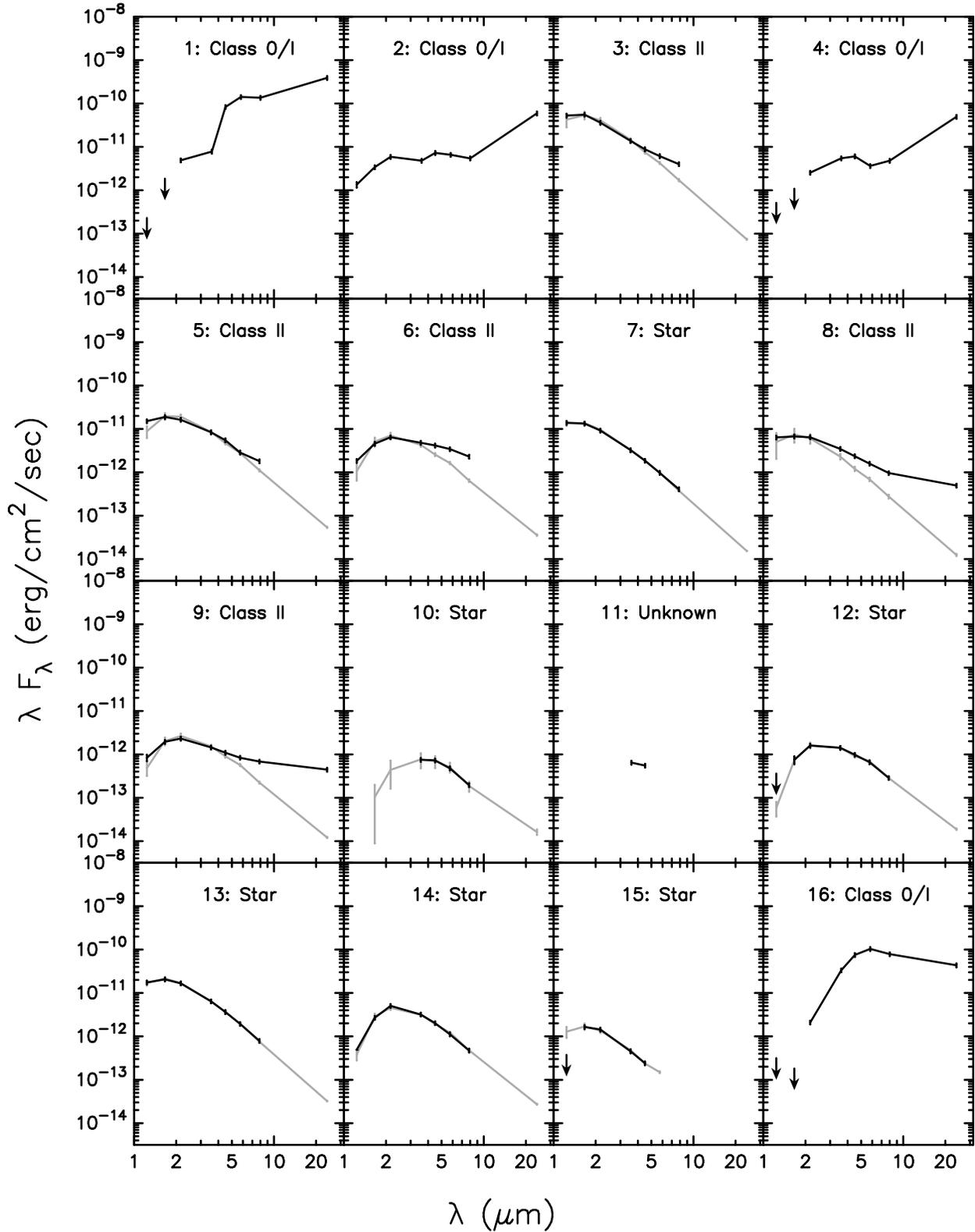}
\figcaption{
The SEDs of sources numbered in Figure 9.  The black lines connect observed 
fluxes while grey lines show respective fits by stellar models.}
\end{figure}

\begin{figure}
\figurenum{15}
\epsscale{0.7}
\plotone{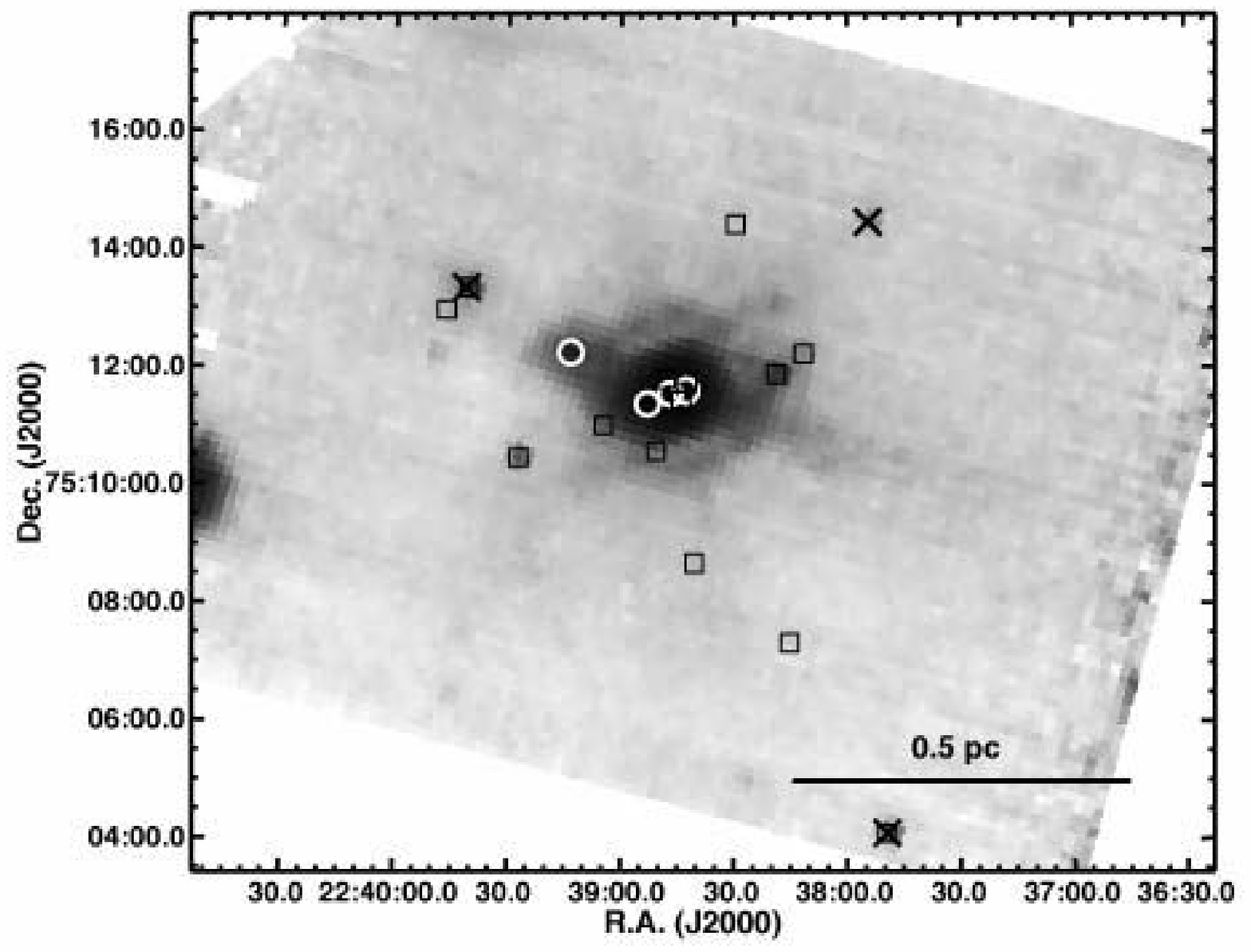}
\epsscale{0.5}
\plotone{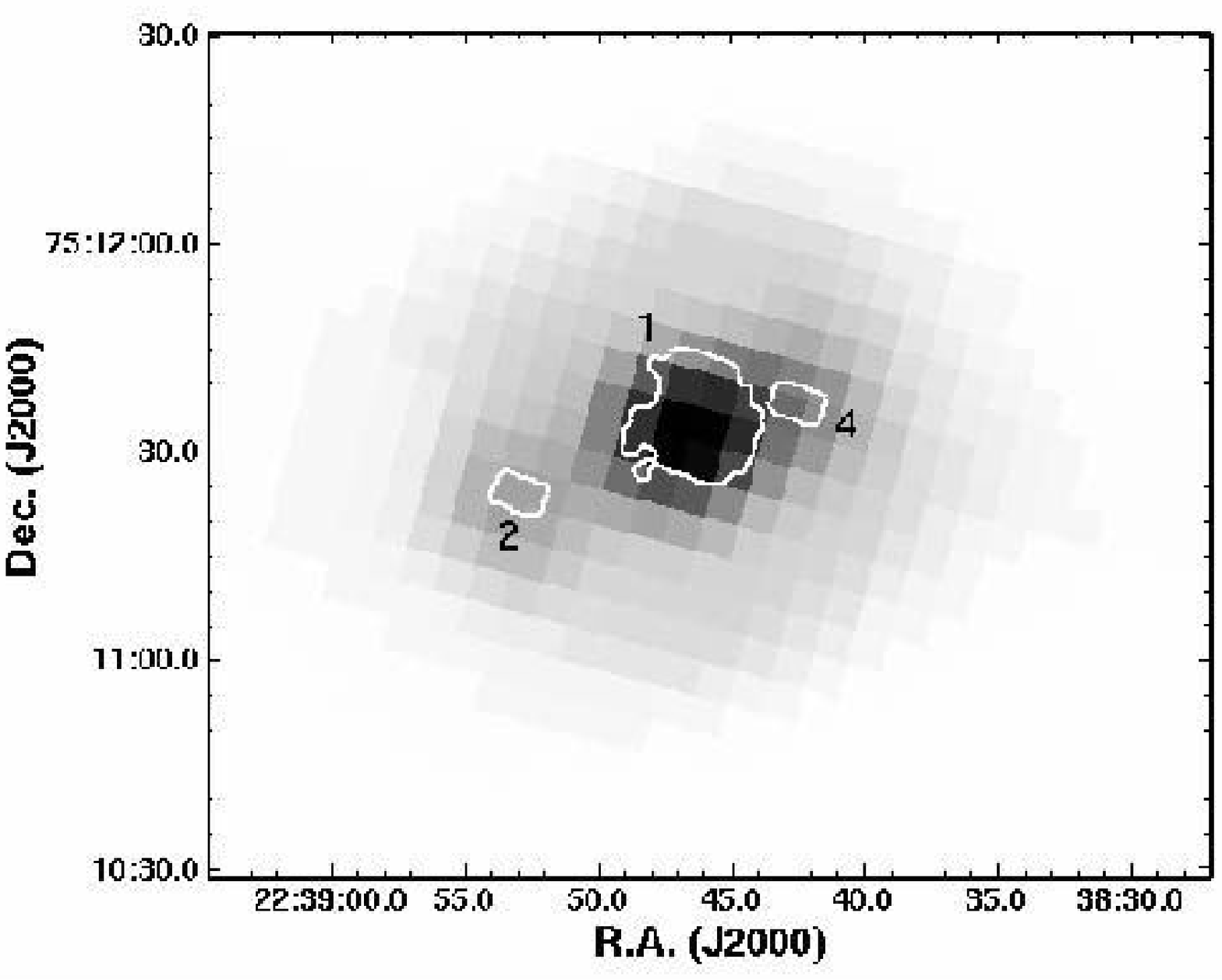}
\figcaption{
a) The L1251E MIPS 70 \micron\ image marked with YSO candidates.  
The symbols are the same as in Figure 12, while X symbols denote the sources 
that are classified as point sources according to the automated Dophot 
photometry output.  The 70 \micron\ fluxes of two Class II candidates, 
which are classified as point sources at 70 \micron, are 122 $(\pm 8.22)$ 
MJy sr$^{-1}$ and 140 $(\pm 5.5)$ MJy sr$^{-1}$ for sources at 
(22$^h$39$^m$41.1$^s$, +75\deg 13\am 21.5\as) and (22$^h$37$^m$49.4$^s$, 
+75\deg 4\am 6.7\as) respectively.  b) The 70 $\mu$m image of the 
L1251B region is shown with 24 $\mu$m contours superposed, indicating the 
precise locations of the three red sources.  The 70 $\mu$m flux measured within 
each contour was used to determine what portions of the total 70 $\mu$m 
flux were divided between IRS1, IRS2 or IRS4.
The diffraction limit is $\sim 7\as$ and $\sim 20\as$ at 24 $\mu$m and 
70 $\mu$m, respectively, for the 85 cm aperture. 
}
\end{figure}

\end{document}